\documentclass[
 jcp,amsmath,amssymb,
 reprint, author-year,
 author-numerical,
]{revtex4-1}

\usepackage{graphicx}
\usepackage{dcolumn}
\usepackage{bm}

\usepackage[utf8]{inputenc}
\usepackage[T1]{fontenc}
\usepackage{mathptmx}
\usepackage{etoolbox}

\usepackage[normalem]{ulem}

\usepackage{chemformula}
\let\ce\ch

\newcommand{\pyrr}{\ce{[pyr14]+}}
\newcommand{\tfsi}{\ce{TFSI$^-$}}
\newcommand{\pyrrtfsi}{\ce{[pyr14][TFSI]}}
\newcommand{\litfsi}{\ce{LiTFSI}}
\newcommand{\li}{\ce{Li$^+$}}
\newcommand{\xli}{x_{\litfsi}}

\usepackage{amsmath}
\usepackage{amssymb}
\usepackage{float}
\restylefloat{table}
\usepackage{longtable}

\usepackage{xr}
\makeatletter
\def\@email#1#2{
 \endgroup
 \patchcmd{\titleblock@produce}
  {\frontmatter@RRAPformat}
  {\frontmatter@RRAPformat{\produce@RRAP{*#1\href{mailto:#2}{#2}}}\frontmatter@RRAPformat}
  {}{}
}
\makeatother
\begin{document}

\preprint{AIP/123-QED}

\title[]{Bridging Electrostatic Screening and Ion Transport in Lithium Salt-Doped Ionic Liquids}
\author{Hyungshick Park$^{a}$, Bong June Sung$^{a,b,\dagger}$, and Jeongmin Kim$^{c,\ast}$}
 \affiliation{$^a$Department of Chemistry, Sogang University, Seoul 04107, Republic of Korea \\
 $^b$Institute of Biological Interfaces, Sogang University, Seoul 04107, Republic of Korea\\
 $^c$Department of Chemistry Education and Graduate Department of Chemical Materials, Pusan National University, Busan 46241, Republic of Korea}
\email{jeongmin@pusan.ac.kr, $^\dagger$bjsung@sogang.ac.kr}

\date{\today}

\begin{abstract}
Alkali salt-doped ionic liquids are emerging as promising electrolyte systems for energy applications, owing to their excellent interfacial stability. To address their limited ionic conductivity, various strategies have been proposed, including modifying the ion solvation environment and enhancing the transport of selected ions (\textit{e.g.}, Li$^+$). Despite the pivotal role of electrostatic interactions in determining key physicochemical properties, their influence on ion transport in such systems has received relatively little attention. In this work, we investigate the connection between ion transport and electrostatic screening using atomistic molecular dynamics simulations of 1-butyl-1-methylpyrrolidinium bis(trifluoromethanesulfonyl)imide ($\pyrrtfsi$) doped with lithium bis(trifluoromethanesulfonyl)imide ($\litfsi$) at molar fractions $\xli\leq0.3$. We find that the charge–charge and density–density correlation functions exhibit oscillatory exponential decay, indicating that $\litfsi$ doped $\pyrrtfsi$ is a charge- and mass-dense system. The electrostatic screening length decreases with increasing $\litfsi$ concentration, whereas the decay length of the density–density correlation functions remains nearly unchanged. Notably, we find that the $\xli$-sensitive screening length serves as a central length scale for disentangling species-specific contributions of ion pairs to collective ion transport upon $\litfsi$ doping. This framework provides a unifying perspective on the interplay between structure and transport in ionic liquid systems.
\end{abstract}

\maketitle

\section{Introduction}
Ionic liquids (ILs) have been extensively studied for a wide range of applications, including green chemistry~\cite{de2022ionic,zhu2024poly}, lubricants~\cite{minami2009ionic,Liu2022}, organic synthesis~\cite{hallett2011room,li2024ionic}, field-effect transistors~\cite{fujimoto2013electric,yadav2021high}, and energy and environmental technologies~\cite{Kalhoff2015,Wilken2015,Roy2022}. Their unique properties, such as negligible volatility, non-flammability, wide electrochemical windows, and high thermal and chemical stability, make them attractive candidates for next-generation functional materials~\cite{Wang2020,Zhou2023}. In particular, their exceptional chemical tunability enables precise tailoring of physicochemical properties for specific target applications~\cite{zeger2025ionic,hu2025thermal,jiang2025research,haddad2025forward}. 

In electrochemical systems, ILs have drawn significant attention as safer alternatives to conventional organic solvents due to their thermal and electrochemical stability~\cite{Wang2020,Crosthwaite2005,xue2018thermal}. However, their practical applications remain limited by relatively low ionic conductivity and high viscosity~\cite{Kubisiak2021,gouverneur2018negative,park2023effects}. To address these challenges, various strategies have been proposed, including dilution with organic solvents, mixing different cations or anions~\cite{chen2019computational,huang2024investigating}, molecular redesign (e.g., ether-functionalized cations, asymmetric anions)\cite{Wettstein2022}, and tuning the salt concentration\cite{lorenz2024evaluating,zhang2024long,park2023simulation}. 
Despite numerous studies, a comprehensive understanding of structure–transport relationships in ionic liquids remains limited, particularly in terms of spatio-temporal correlations. 

From recent surface force balance (SFB) experiments~\cite{Gebbie2015,Jager2023,Lee2017a}, electrostatic screening behaviors, one of structural features, of ionic liquids have drawn significant attention. These studies revealed the ILs, not simply deviating from the Debye-H\"uckel theory, exhibit unexpectedly long electrostatic decay lengths ($\lambda_s$), reaching up to $\sim$10 nm in some cases. Such anomalous \textit{underscreening}, where $\lambda_s$ exceeds the Debye screening length and increases with salt concentration, emerged with the longer $\lambda_s$ in more concentrated systems, across a broad range of concentrated electrolytes, including polymerized ILs and alkali metal salt-doped ILs~\cite{Avni2020,Goodwin2017,Lee2017,Krucker-Velasquez2021,Rotenberg2018,zhang2024long}. The SFB studies identified a transition from a dilute to a concentrated regime when $\lambda_s$ becomes comparable to the ion size, with underscreening observed primarily in the concentrated regime. Such a transition was previously predicted by Kirkwood~\cite{kirkwood1936statistical}. Notably, in the concentrated regime, a universal scaling relation was found across chemically diverse ionic systems by normalizing $\lambda_s$ and the mean ion size $a$ with the Debye screening length $\lambda_D$ as follows:
\begin{equation}
\bigg(\frac{\lambda_s}{\lambda_D}\bigg)\sim\bigg(\frac{a}{\lambda_D}\bigg)^{\alpha},
\end{equation}
where the scaling exponent $\alpha = 3$ in the SFB experiments. Bulk molecular simulations~\cite{coles2020correlation,zeman2021ionic,Krucker-Velasquez2021} and classical density functional theories~\cite{Adar2019,Cats2021,Rotenberg2018} have also captured a similar transition and universal scaling behavior, although the exponent $\alpha$ was reported to be lower, ranging from 1 to 2. Despite numerous studies including the ion cluster theory, the phenomenon of underscreening remains not fully understood; some studies~\cite{hartel2023anomalous,Kumar2022,wang2024structure} recently suggested that confinement- and interface-induced effects, such as pronounced ion association, could explain the cubic scaling observed in SFB experiments.

Introducing lithium salts into ILs introduces additional structural and dynamical complexities~\cite{Kubisiak2021,Roy2022,Giffin2017,Zhong2023,Thum2020,Tong2020,Pitawala2012,Francis2022,Haskins2014,Maiti2023,park2022effects}. The alkali metal ions, due to the strong electrostatic interactions, significantly influence both the coordination environment and conformational preferences of the constituent ions, such as the denticity of TFSI anions~\cite{li2012li+,Haskins2014,Lesch2016,Wettstein2022,c2021molecular,Tong2020}. Beyond these local effects, lithium salts also induce the formation of ion clusters composed of \li~and coordinating anions. These clusters are often asymmetric and negatively charged, as a single \li~can coordinate with multiple anions, and vice versa. Such strong static correlations lead to dynamic correlations, exemplified by the emergence of negative lithium transference numbers ($t_{\ce{Li}}$) in pulsed-field gradient NMR (PFG-NMR) measurements~\cite{Kubisiak2021,gouverneur2018negative}. This counterintuitive phenomenon is typically attributed to the vehicular motion of negatively charged clusters (e.g., \tfsi–\li–\tfsi), which collectively transport lithium ions in the “wrong” direction~\cite{Molinari2019,Molinari2019a,Dong2018,McEldrew2021b}.
Several studies have aimed to correlate the formation of Li–TFSI clusters to the ion transport. The approaches, addressing a central question being whether these clusters diffuse via a vehicular or structural mechanism, include the comparison between two length scales of diffusion and solvation shell dynamics~\cite{Self2019},  the lithium coupling factor (LCF), which suggested the flow-like coupled diffusion~\cite{Wettstein2022}, and ion cluster theory~\cite{McEldrew2021,Goodwin2023,Kornyshev2022, mceldrew2020theory,zhang2024long}. 

In this work, building on these insights, we demonstrate that the electrostatic screening length, $\lambda_Z$, serves as a central and physically meaningful length scale for characterizing such correlated ion motion in the lithium-doped ILs (LILs) using atomistic molecular dynamics simulations. Our approach offers a unifying perspective on the interplay between structure and transport in LILs. We investigate mixtures of 1-butyl-1-methylpyrrolidinium bis(trifluoromethanesulfonyl)imide (\pyrrtfsi) and lithium bis(trifluoromethanesulfonyl)imide (\litfsi) at molar fractions $\xli = 0$, 0.1, 0.2, and 0.3. $\pyrrtfsi$ is a widely studied IL system due to its potential in practical applications such as lithium-ion batteries~\cite{balducci2018ionic,Kalhoff2015}. Moreover, pure \pyrrtfsi \textcolor{red}{~}exhibits anomalous underscreening with a long-range surface force decay (up to $\sim$8.4 nm) as measured via surface force balance (SFB) experiments~\cite{smith2016electrostatic}. In this prototypical LIL system, we explore how lithium salt doping alters both short-range structures and long-range charge correlations. We find that the charge–charge and density–density correlation functions exhibit oscillatory exponential decay, indicating that LILs are charge- and mass-dense systems. Interestingly, while the electrostatic screening length decreases with increasing \litfsi\textcolor{red}{~}concentration, the decay length of the density–density correlations remains nearly unchanged. Notably, we find that the $\xli$-dependent screening length serves as a central length scale for disentangling species-specific contributions to collective ion transport. Our analysis further suggests that increased Li–TFSI associations at higher $\xli$ concentrations promote the formation of ionic clusters, which in turn liberate $\pyrr$ cations and enhance the overall \textit{ionicity} of the LILs. 

The rest of the paper is organized as follows. In Section~\ref{sec:method}, we present the simulation model and force fields employed in this study, detailing the calculation of both static properties (\emph{e.g.}, charge-charge and density-density correlation functions) and dynamic properties (\emph{e.g.}, ionic conductivity, and transference number). In Section~\ref{sec:results}, we first examine the impact of $\li$ doping on the decay of structural correlations at long distances. We then explore how the screening length can serve as a bridge to understanding correlated ion transport in LILs, with a particular focus on species-dependent diffusion behavior. Finally, Section~\ref{sec:conclusions} provides a summary of this work and discusses potential future directions.

\begin {figure}[t]
\centering
\includegraphics[width=3.5in]{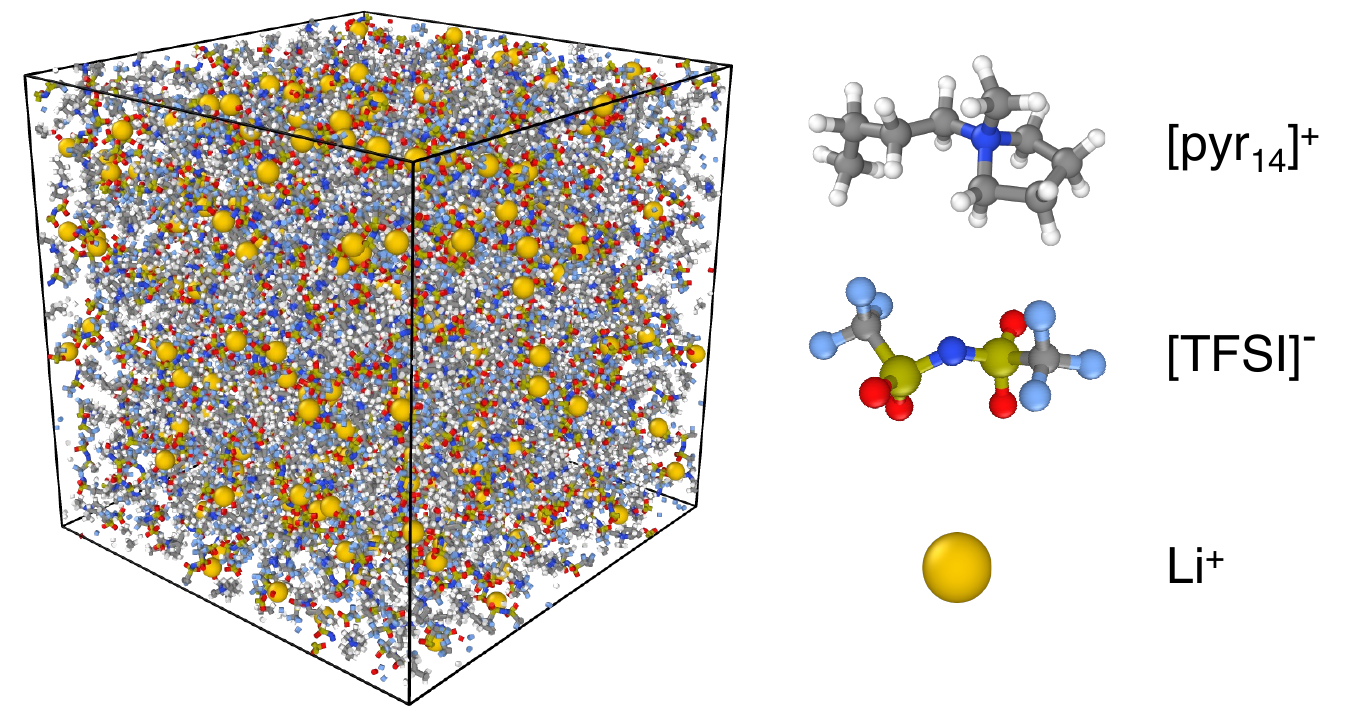}
\caption{Simulation snapshot of $\litfsi$-doped $\pyrrtfsi$ at a  $\litfsi$ molar fraction of $\xli=0.3$. Lithium ions are shown as yellow spheres, while $\pyrr$ and $\tfsi$ ions are depicted as stick models.}
\label{fig:snap}
\end{figure}

\section{Simulation Model and Methods}\label{sec:method}
We conducted atomistic molecular dynamics simulations of LILs with varying salt concentration to investigate the effect of $\li$ doping on the charge screening and transport properties of LILs. In this section, we introduce the target IL system (Fig~\ref{fig:snap}), force fields, simulation details and computing the observables.

\subsection{Simulation model: LiTFSI-doped $\pyrrtfsi$}
We model the pure and lithium salt-doped $\pyrrtfsi$ (Fig.~\ref{fig:snap}) using the widely-used CL\&P force field developed by P\'{a}dua and co-workers~\cite{shimizu2010molecular,Lopes2004,Voroshylova2014,Chaban2015,Goloviznina2019}. The functional form, identical to the conventional OPLS-AA force field, of the total energy $V_{tot}$ is as follows: 
\begin{eqnarray}
V_{tot} &=& \sum_{bonds} k_b(r-r_0)^2 +  \sum_{angles} k_\theta(\theta-\theta_0)^2\\ \nonumber
              &+&  \sum_{dihedrals}\sum_{m=1}^{n} \frac{k_\chi}{2}[1+(-1)^{m+1}\cos(m\chi)]\\ \nonumber
              &+&  \sum_{impropers}\sum_{m=1}^{n} \frac{k_\psi}{2}[1+(-1)^{m+1}\cos(m\psi)]\\ \nonumber
              &+& \sum_{i}\sum_{i<j}\left\{4\epsilon_{ij} \left[\left(\frac{\sigma_{ij}}{r_{ij}} \right)^{12}-\left(\frac{\sigma_{ij}}{r_{ij}} \right)^6\right] + \frac{q_iq_j}{4\pi\varepsilon_0r_{ij}}\right \}.
\label{potential}
\end{eqnarray}
The first four terms describe the intramolecular interactions, while the last term is for the non-bonding interactions, consisting of Lenard-Jones and Coulomb interactions with the particle indices $i$ and $j$. Here, $k_b$, $k_\theta$, $k_\chi$, and $k_\psi$ denote the bond, bond angle, dihedral, and improper interaction coefficients, respectively. $r_0$ and $\theta_0$ are the bond distance and bond angle at energy minimum, respectively. The non-bonding interactions consist of Lennard-Jones and Coulomb interactions between $i^{th}$ and $j^{th}$ particles. 

Both non-bonding interactions were cut at 1.2 nm, and the long-range part of the Coulomb interaction was calculated by using the Particle Mesh Ewald~\cite{essmann1995smooth}. We should note that the 0.8 rescaled charges were used for the Coulomb interactions, while the full charges were accounted for to estimate the thermodynamic and dynamic properties of the systems~\cite{blazquez2023computation}. 
In other words, scaled charges were used to construct the potential energy surface (PES) and evolve the system, while full charges were used during post-processing to compute the ionic conductivity. This is because transport properties that involve response to external fields, such as ionic conductivity, depend on both the dipole moment surface and the PES~\cite{blazquez2023computation}. This two-surface approach has previously yielded excellent agreement with the experimental conductivities of aqueous NaCl and KCl solutions up to their solubility limits.
The force field parameters are obtained from \ce{fftool} developed by P\'{a}dua group~\cite{agilio_padua_2021_4701065}. All the systems consist of a total of 1000 pairs of cations and anions, and the fraction of $\li$ ions ($\xli$) among a total of 1000 cations ranges from 0 to 0.3.

\subsection{Simulation method.}
We carried out all-atom molecular dynamics simulations in the isobaric-isothermal ($NpT$) ensemble by using OpenMM with CUDA acceleration~\cite{eastman2017openmm}. Initial coordinates were generated via Packmol package~\cite{martinez2009packmol} and the initial random velocities are generated to follow the Maxwell distribution at 400K. During the equilibration of 300 ns, we integrated the equations of motion using the Langevin integrator at 400 K with friction coefficient 1 ps$^{-1}$, after which both density and total energy well converge. In the production run, we employed simulations with the Nos\'{e}-Hoover integrator at 400 K with the collision frequency 1 ps$^{-1}$. In both equilibration and production runs, the desired pressure ($p=1$ atm) was maintained using the isotropic Monte Carlo barostat implemented in OpenMM~\cite{eastman2017openmm}. The observables were computed using five independent trajectories of at least 300 ns; we performed longer production simulations (up to 1,300 ns) due to the slow dynamics at high concentrations of $\litfsi$. 

\subsection{Computation of observables}

In this work, we investigated both the structural and dynamical properties of the LILs. To investigate the structural properties, we analyzed the number-number or charge-charge correlation functions, and local charge atmosphere. For the dynamic properties, we analyzed the ionic conductivity, transference number, and ion pair dynamics.

\textbf{The charge-charge and number-number correlation functions.} To quantify the long-range decay of the ion structures, we calculated the number-number ($g_{NN}(r)$) and charge-charge ($g_{ZZ}(r)$) correlation function as follows,
\begin{eqnarray}
g_{NN}(r) = \sum_{\alpha} \sum_{\beta}  \chi_{\alpha} \chi_{\beta} g_{\alpha\beta}(r),
\label{eq:gnn}
\end{eqnarray}
\begin{eqnarray}
g_{ZZ}(r) = \sum_{\alpha} \sum_{\beta}  Z_{\alpha} Z_{\beta} \chi_{\alpha} \chi_{\beta} g_{\alpha\beta}(r), 
\label{eq:gzz}
\end{eqnarray}
where $Z_{\alpha}$ and $\chi_{\alpha}$($=N_\alpha/N$) denote the charge and number fraction of the ionic species $\alpha$, respectively. The electro-neutrality ensures that $\sum_{\alpha} Z_{\alpha}\chi_{\alpha} =0$. Here, $g_{\alpha\beta}$ indicates the radial distribution function (RDF) between ionic species $\alpha$ and $\beta$.
\begin{eqnarray}
g_{\alpha\beta}(r) = \frac{V}{4 \pi r^2 N_{\alpha} N_{\beta}} \sum_{i\neq j} \langle \delta(r-|\vec{r}^{\alpha}_i - \vec{r}^{\beta}_j|) \rangle,
\label{eq:gr}
\end{eqnarray}
where $V$, and $N_{\alpha}$ denote the volume of a simulation cell, the number of species $\alpha$, respectively. In this work, we used a different definition of $\vec{r}^{\alpha}_i$: for $g_{ZZ}(r)$, it denotes the center of charge of the $i^{\text{th}}$ ion of species $\alpha$, whereas for $g_{NN}(r)$, it refers to the center of mass of the same ion. $\langle \cdots \rangle$ denotes the ensemble average. In the ionic liquids and concentrated electrolytes, the decay of the correlation functions often follow an oscillatory exponential function as below~\cite{coles2020correlation, zeman2021ionic, keblinski2000molecular}: 
\begin{eqnarray}
\ln(|g_{ZZ}(r)\cdot r|) = - \frac{r}{\lambda_{Z}} + \ln\bigg[ A_Z \bigg| \cos\bigg(\frac{2\pi r}{l_{Z}} + \phi_{Z}\bigg) \bigg| \bigg].
\label{eq:gzz_fit}
\end{eqnarray}
The first term represents the exponential decay with the decay length, $\lambda_Z$, and the second term represents the oscillation with $A_Z$, $\phi_Z$, and $l_Z$ being its amplitude, phase, and period, respectively. As opposed to concentrated electrolytes as in our case, there appears no oscillation for sufficiently dilute electrolytes, following the Debye-H\"uckel theory~\cite{debye1923theory,avni2022conductivity,shilov2015role}. 
Similarly, the decay of the $g_{NN}(r)$ can be analyzed with the oscillatory exponential function.
\begin{eqnarray}
\ln(|(g_{NN}(r)-1)\cdot r|) = - \frac{r}{\lambda_{N}} + \ln\bigg[ A_N \bigg| \cos\bigg(\frac{2\pi r}{l_{N}} + \phi_{N}\bigg) \bigg| \bigg].
\label{eq:gnn_fit}
\end{eqnarray}
We note that $g_{NN}(r\rightarrow\infty)\rightarrow1$, while $g_{ZZ}(r\rightarrow\infty)\rightarrow0$. To extract physical quantities, we fitted both correlation functions with Eq.~\ref{eq:gzz_fit} and~\ref{eq:gnn_fit} within the ranges of 1 nm$~<r<$~3.5 nm and 0.5 nm$~<r<$~2.5 nm, respectively (see Table~\ref{si:tab:gzzgnn} for the values). 
In this work, we utilized Eqs.~\ref{eq:gzz_fit} and~\ref{eq:gnn_fit} under the assumption that a \textit{single} Yukawa-type function can describe the decay of $g_{ZZ}(r)$ and $g_{NN}(r)$. In general, however, multiple decay modes can coexist, as described in the dressed ion theory~\cite{kjellander2019intimate,kjellander2020multiple}. For instance, recent simulation studies have shown that these correlation functions are better described by a sum of two or three Yukawa-type functions~\cite{zeman2020bulk,hartel2023anomalous}. Nevertheless, despite the simplification, the extracted decay lengths using a single decay mode provide a consistent and practical framework for comparing electrostatic screening behavior across different compositions.

\textbf{Local ion atmosphere.} 
To investigate the local charge screening around a specific ion, we computed the cumulative charge distribution ($c^Q_{\alpha}(r)$) around the ionic species $\alpha$ using the RDFs. 
\begin{eqnarray}\label{eq:cQr}
c^{Q}_{\alpha}(r) &=& Z_{\alpha} + \sum_{\beta} Z_{\beta} c_{\alpha\beta}(r), \\
&=&Z_{\alpha} +\sum_{\beta} Z_{\beta}\frac{N_\beta}{V}\int_0^{r} 4\pi r^{\prime 2} g_{\alpha\beta}(r^{\prime})dr^{\prime}.
\end{eqnarray}
Here, $c_{\alpha\beta}(r)$ is a cumulative distribution of species $\beta$ around species $\alpha$. It then satisfies the relations that $c^{Q}_{\alpha}(r=0)=Z_{\alpha}$, and  $c^{Q}_{\alpha}(r\rightarrow\infty)\rightarrow0$. 
In the LILs investigated, $c^{Q}_{\alpha}(r)$ show oscillation as for $g_{ZZ}(r)$ due to the significant steric interactions between the ions. One can obtain an analytical expression for a long-range decay of $c^{Q}_{\alpha}(r)$, when assuming that $g_{\alpha\beta}(r)$ follows a similar oscillatory decay to $g_{ZZ}(r)$ (Eq.~\ref{eq:gzz_fit}) at large enough distances.
\begin{equation}\begin{split}\label{eq:cq_fit}
c^{Q}_{\alpha}(r) & \approx Z_\alpha -\frac{4\pi A^{Q}_{\alpha} l_Z \lambda_Z}{(l_Z^2+4\pi^2\lambda_Z^2)^2} \bigg[ l_Z \left( l_Z^2 \left( \lambda_Z + r \right) - 4\pi^2 \lambda_Z^2 \left( \lambda_Z - r \right) \right) \\ 
& \text{cos}\left(\phi^{Q}_{\alpha} + \frac{2\pi r}{l_Z} \right)  - 2 \pi \lambda_Z \left( l_Z^2 \left(2 \lambda_Z + r \right) + 4 \pi^2 \lambda_Z^2 r \right) \\ 
& \text{sin} \left(\phi^{Q}_{\alpha} + \frac{2\pi r}{l_Z} \right) \bigg] \exp\left(-\frac{r}{\lambda_Z}\right).
\end{split}\end{equation}
Here, we assumed the species-\textit{independent} decay length ($\lambda_Z$) and the oscillation period ($l_Z$), while the amplitude ($A^{Q}_{\alpha}$) and ($\phi^{Q}_{\alpha}$) are species-specific. Our assumption is in line with the one in Lee et al.~\cite{Lee2017a} that in the asymptotic limit all correlation functions decay at the same rate and with the same oscillation period. We note that in case that $\phi^{Q}_{\alpha}\rightarrow0$ and $l_Z\rightarrow\infty$, $c^{Q}_{\alpha}(r)$ becomes a simple exponential function, as expected from the Debye-H\"uckel theory (see Section~\ref{si:sec:cQr_DH} in the SI for derivation). 
The oscillatory decay of $c^Q (r)$ in Eq.~\ref{eq:cq_fit} is often referred to as \textit{overscreening}~\cite{Feng2019, begic2019overscreening,bazant2011double}, which occurs when surrounding counterions overcompensate the charge of a central ion. This behavior stems from the damped oscillations in $g_{ZZ}(r)$ as described in Eq.~\ref{eq:gzz_fit}. In the present work, we use the term \textit{underscreening} to describe cases where the electrostatic screening length is longer than the Debye screening length. We note that these two effects, underscreening and overscreening, are not mutually exclusive~\cite{Lee2017a} and are found to coexist in our LIL systems. Indeed, simulation results~\cite{zeman2021ionic,hartel2023anomalous} often report oscillatory decay in $g_{ZZ}(r)$ with a screening length longer than predicted by the Debye–Hückel theory, consistent with our observations.

\begin {figure*}[thb]
\centering
\includegraphics[width=5.in]{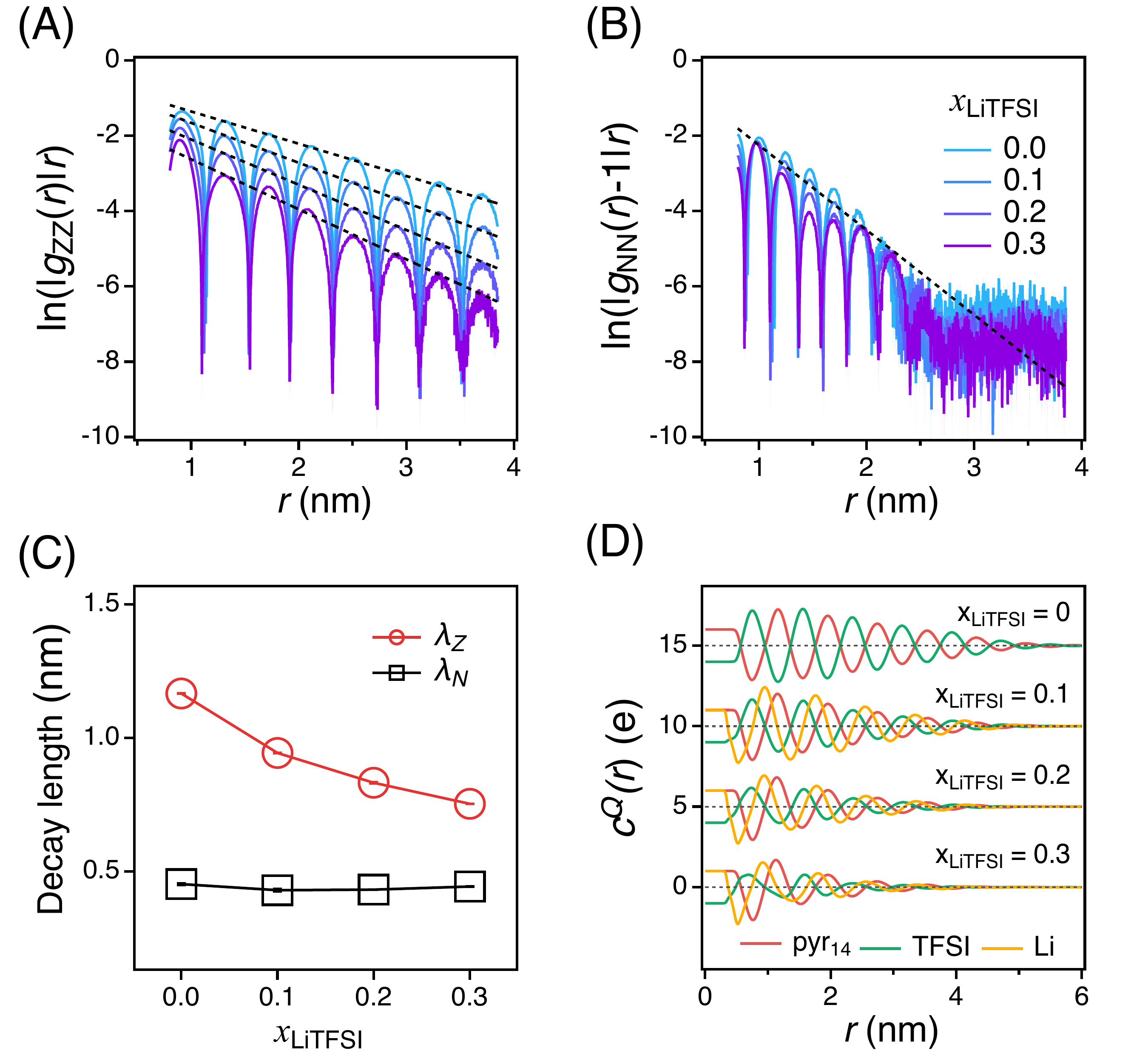}
\caption{Structural correlations at various $\litfsi$ doping fractions, $\xli$. (A) Charge-charge correlation function ($g_{ZZ}(r)$, Eq.~\ref{eq:gzz}). The dashed lines represent the fits to the exponential part of Eq.~\ref{eq:gzz_fit} with the decay length $\lambda_Z$. (B) Density-density correlation function ($g_{NN}(r)$, Eq.~\ref{eq:gnn}). The dashed line represents the fits to the exponential part of Eq.~\ref{eq:gnn_fit} with the decay length $\lambda_N$ at $\xli=0.3$. (C) Extracted decay lengths $\lambda_Z$ and $\lambda_N$ as a function of $\xli$. (D) Ion atmosphere profiles (Eq.~\ref{eq:cQr}) of the constituent ions. The curves at each $\xli$ are vertically shifted for visual clarity. See Table~\ref{si:tab:ionatm} for fitting results.} 
\label{fig:gzzgnn}
\end{figure*}

\textbf{Ionic conductivity and correlated motion.} 
We calculated Onsager transport coefficients to estimate the ionic conductivity, accounting for all contributions of the correlated motions~\cite{fong2020,fong2021ion,wheeler2004molecular,Molinari2019a}. The mean-squared charge displacement $\Sigma$ was computed as follows,
\begin{eqnarray}
\Sigma(t) &=& \frac{1}{Vk_BT} \sum_{\alpha}\sum_{\beta} {Z_\alpha Z_\beta} \Delta{r}_{\alpha\beta}(t) \\
&=& \frac{1}{Vk_BT} \sum_{\alpha}\sum_{\beta} {Z_\alpha Z_\beta} \sum^{N_\alpha}_{i\in\alpha} \sum^{N_\beta}_{j\in\beta}  \Delta {r}^{\alpha\beta}_{ij}(t),
\label{eq:og_component}
\end{eqnarray}
where $k_B$ and $T$ denote the Boltzmann constant and temperature, respectively. 
Here, $\Delta {r}^{\alpha\beta}_{ij}$ is the displacements product:
\begin{eqnarray}
\Delta {r}^{\alpha\beta}_{ij} \equiv \langle \Delta \vec{r}_i^{\alpha}(t) \cdot \Delta \vec{r}_j^{\beta}(t) \rangle,
\end{eqnarray}
where $\Delta \vec{r}_i^{\alpha}(t)~(=\vec{r}_i^{\alpha}(t)- \vec{r}_i^{\alpha}(0))$ denotes the displacement of the $i^{th}$ ion of species $\alpha$ during time $t$. In the long-time limit, $\Sigma(t)$ grows linearly with time $t$ and its slope is the ionic conductivity, $\sigma$:
\begin{eqnarray}
\sigma= \lim_{t\to\infty} \frac{\Sigma(t)}{6t},
\label{eq:gk_slope}
\end{eqnarray}
One can then decompose the contributions to $\sigma$ as follows (see Sections~\ref{si:sec:diff} and~\ref{si:sec:cond} in the SI for further details): 
\begin{eqnarray}
\sigma &=& \sigma_{\ce{pyr_{14}}}^s + \sigma_{\ce{TFSI}}^s + \sigma_{\ce{Li}}^s + \sigma_{\ce{pyr_{14}}}^d + \sigma_{\ce{TFSI}}^d + \sigma_{\ce{Li}}^d \\
            &+& 2\sigma_{\ce{pyr_{14}}-\ce{TFSI}} + 2\sigma_{\ce{Li}-\ce{pyr_{14}}}+ 2\sigma_{\ce{Li}-\ce{TFSI}}.
\label{eq:gk_conductivity}
\end{eqnarray}
Here, the superscripts $s$ and $d$ denote the self and distinct contributions, respectively. The subscripts indicates the ion species that contribute to $\sigma$. One can also compute the lithium transference number $t_{\ce{Li}}$ to estimate the fraction of ionic current delivered by $\li$: 
\begin{eqnarray}\label{eq:tLi}
t_{\ce{Li}} \equiv \frac{\sigma_{\ce{Li}}^s + \sigma_{\ce{Li}}^d + \sigma_{\ce{Li}-\ce{pyr_{14}}} + \sigma_{\ce{Li}-\ce{TFSI}}}{\sigma}. 
\end{eqnarray}
Similarly, the transference number can be defined for all constituent ions, satisfying $\sum_{\alpha} t_{\alpha} = 1$ with $\alpha\in$ \{Li, pyr14, TFSI\}.

\textbf{Nernst-Einstein approximation.} By neglecting all ion–ion correlations, the ionic conductivity can be approximated using the Nernst–Einstein (NE) relationship:
 \begin{eqnarray}\label{eq:ne_conductivity}
\sigma_{NE}&=&\sigma_{\ce{pyr_{14}}}^s + \sigma_{\ce{TFSI}}^s + \sigma_{\ce{Li}}^s\nonumber\\
&=&\frac{\rho_{\text{ion}}e^2}{k_BT}\bigg(\chi_{\ce{pyr_{14}}}D_{\ce{pyr_{14}}}+\chi_{\ce{TFSI}}D_{\ce{TFSI}}+\chi_{\ce{Li}}D_{\ce{Li}}\bigg).
\end{eqnarray}

The contribution $\sigma_{corr}$ of correlated motions to the ionic conductivity is $\sigma_{corr}=\sigma-\sigma_{NE}$ and $\sigma/\sigma_{NE}=1+\sigma_{corr}/\sigma_{NE}$. $\sigma_{corr}$ is usually negative as the dynamic correlations often decrease $\sigma$. Under the NE approximation, 
 \begin{eqnarray}\nonumber
 t^{NE}_{\ce{Li}} &=& \sigma_{\ce{Li}}^s/\sigma_{NE} \\ \nonumber
 &=&(N_{\ce{Li}}D_{\ce{Li}})/\big(N_{\ce{pyr_{14}}}D_{\ce{pyr_{14}}}+N_{\ce{TFSI}}D_{\ce{TFSI}}+N_{\ce{Li}}D_{\ce{Li}}\big).
 \end{eqnarray} 
 Again, this satisfies that $\sum_{\alpha} t^{NE}_{\alpha} = 1$ with $\alpha\in$ \{Li, pyr$_{14}$, TFSI\}. 

\textbf{Ion pair dynamics.} To quantify the lifetime of ion pairs, we computed the following time correlation function:
\begin{equation}\label{eq:pairdyn}
    H_{\alpha\beta}(t) = \langle h_{ij}(t) h_{ij}(0) \rangle,
\end{equation}
where $h_{ij}(t)$ is an indicator function defined such that $h_{ij}(t) = 1$ if ions $i$ and $j$ form a neighboring pair within a certain distance at time $t$, and 0 otherwise. Here, ions $i$ and $j$ belong to species $\alpha$ and $\beta$, respectively. For each $\alpha$–$\beta$ pair, $H_{\alpha\beta}(t)$ decays from 1 at $t = 0$ to 0 as $t \to \infty$. In this work, we used several distance-based criteria ($L_s$) to define neighboring ion pairs at each time, including $\xli$-dependent $\lambda_Z$, $\lambda_Z$ of pure $\pyrrtfsi$, and the first minimum, $r_{min}$, of $g_{\alpha\beta}(r)$. To quantify the decay of $H_{\alpha\beta}(t)$ we fit it to a stretched exponential function: $H_{\alpha\beta}(t)\approx \exp[-(t/\tau_H)^{\beta_H}]$. where 
$\tau_H$ is a characteristic time constant, and $\beta_H$ is the stretching exponent. The mean lifetime is then estimated as $\langle\tau_H\rangle=\int_0^\infty tH_{\alpha\beta}(t)dt=\tau_{\text{H}}\Gamma(1/\beta_{\text{H}})/\beta_{\text{H}}$, where $\Gamma$ denotes the Gamma function.

\section{Results and Discussion}\label{sec:results}

In this section, we discuss simulation results of short- and long-ranged structures, screening lengths ($\lambda_Z$ and $\lambda_N$) in Li-doped $\pyrrtfsi$ at various $\xli$. In addition, we estimate the extent of correlative motions, ionic conductivities and $\li$ transference number, and discuss about their relations with charge screening in our system.

The primary effect of $\li$ doping is condensation, driven by the high charge density and small size of $\li$ ions compared to the larger organic ions. The number density of ions increases from 3.0~M to 3.6~M with increasing $\xli$, reflecting the condensed ionic structures and increased densities (see Table~\ref{si:tab:props} in the SI). Additionally, the solvation structure of $\li$ evolves with varying $\xli$. $\tfsi$ ions preferentially solvate $\li$ due to strong electrostatic interactions. With increasing $\xli$, the conformational structure of solvating $\tfsi$ ions shifts toward the \textit{cis} conformation~\cite{li2012li+,Haskins2014,Lesch2016,Wettstein2022,c2021molecular}. Furthermore, monodentate coordination of $\li$ becomes more prevalent than bidentate coordination~\cite{Haskins2014, Wettstein2022, fujii2006conformational, c2021molecular}. Results of these short-range structural changes are given in the SI (Figs.~\ref{si:fig:rdf} and~\ref{si:fig:Li_coord}). In this work, we investigate how lithium salt doping affects the decay of charge–charge ($g_{ZZ}(r)$) and density–density ($g_{NN}(r)$) correlation functions in ILs. Our analysis reveals structural changes that extend beyond the condensation and short-range solvation environment, focusing on longer-range correlations that provide connections to collective ionic transport.

\subsection{Screening length decreases with increasing lithium salt fraction in charge- and mass-dense LILs}\label{sec:screening}

\begin {figure}[]
\centering
\includegraphics[width=2.7in]{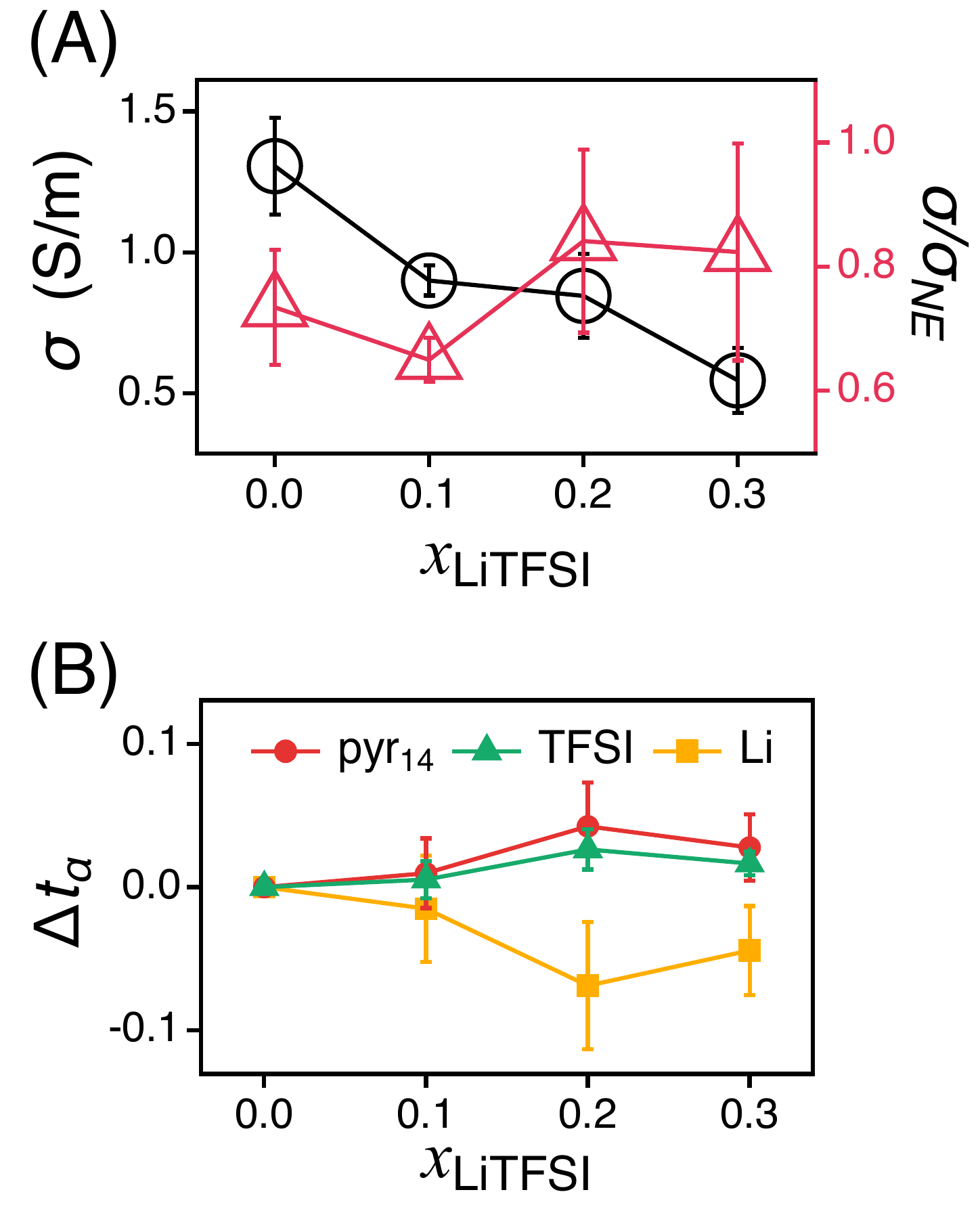}
\caption{Simulation results of ion transport at various $\litfsi$ doping fractions, $\xli$. (A) Ionic conductivity ($\sigma$, Eq.~\ref{eq:gk_conductivity}) and its ratio to the approximate value $\sigma_{NE}$ from the Nernst-Einstein relation (Eq.~\ref{eq:ne_conductivity}). (B) Change in transference number, $\Delta t_\alpha=t_\alpha-t_\alpha(\xli=0)$, for ion species $\alpha\in\{\ce{Li},\ce{pyr_{14}},\ce{TFSI}\}$. See Fig.~\ref{si:fig:cond} in the SI for $t_\alpha$.} 
\label{fig:cond}
\end{figure}

Figures.~\ref{fig:gzzgnn}A and~\ref{fig:gzzgnn}B demonstrate that both $g_{ZZ}(r)$ and $g_{NN}(r)$ exhibit features of oscillatory exponential decay in both pure and $\litfsi$-doped $\pyrrtfsi$, indicating that these systems are mass- and charge-dense~\cite{bazant2011double, coles2020correlation, Coupette2018, keblinski2000molecular, shim2005molecular, hartel2023anomalous}. The oscillatory exponential behavior clearly deviates from the conventional Debye-H\"uckel theory, suggesting the significant contributions of other molecular forces, \emph{e.g.}, steric interactions, in such dense Coulomb systems~\cite{keblinski2000molecular, shim2005molecular,zeman2021ionic}. The observed values of $\lambda_Z$ are similar to the ones reported by previous simulation studies, while shorter than the ones by SFB experiments~\cite{Gebbie2015,Jager2023,Lee2017a, zeman2021ionic}. As shown in Fig.~\ref{fig:gzzgnn}C, in pure $\pyrrtfsi$, the decay length of $g_{ZZ}(r)$ is as approximately twice long as that of $g_{NN}(r)$, and the oscillation period of $g_{ZZ}(r)$ is longer than that of $g_{NN}(r)$ (see Table~\ref{si:tab:gzzgnn} in the SI). Upon $\litfsi$ doping, a distinct response emerges in the decay behavior of the two correlation functions. The oscillation periods of both $g_{ZZ}(r)$ and $g_{NN}(r)$ remain essentially unchanged, maintaining values similar to those in pure $\pyrrtfsi$. However, the decay lengths show contrasting trends with increasing $\xli$: $\lambda_Z$ associated with $g_{ZZ}(r)$ decreases significantly, indicating faster decay, whereas $\lambda_N$ of $g_{NN}(r)$ remains relatively constant (Fig.~\ref{fig:gzzgnn}C). Notably, over the investigated $\xli$ range, $\lambda_Z$ consistently exceeds both $\lambda_N$ and the molecular ion size. 
At the same time, with increasing $\xli$, the charge-number coupling is enhanced, with $\lambda_Z$ approaching $\lambda_N$. Owing to the significant size asymmetry between Li$^+$ and the other molecular ions, the charge-number correlation $g_{ZN}(r)$ becomes more pronounced, and its decay length becomes comparable to $\lambda_Z$ at $\xli$ = 0.3~\cite{kjellander2019intimate,kjellander2020multiple} (see Fig.~\ref{si:fig:corr} in the SI). Despite the enhanced charge-number coupling, a single decay mode remained sufficient to describe all three correlation functions ($g_{ZZ}(r)$, $g_{NN}(r)$, and $g_{ZN}(r)$) at $\xli$ = 0.3 (see Fig.~\ref{si:fig:fit} in the SI). Further investigation could elucidate the effects of lithium salt doping in ionic liquids on the charge-number coupling, driven by pronounced size asymmetry, which is beyond the scope of the present study. Thus, such coupling and decoupling highlight the complex interplay between mass and charge organization in these systems. 

The features of the correlation functions (Figs.~\ref{fig:gzzgnn}A and~\ref{fig:gzzgnn}B) are closely related to the structural organization of ionic liquid systems and its evolution with lithium doping, which manifests in the corresponding structure factors~\cite{amith2020pictorial,umebayashi2011liquid,aguilera2015effect} and surface force profiles~\cite{miao2024ion}. The presence of oscillatory decays in $g_{ZZ}(r)$ and $g_{NN}(r)$ implies the existence of corresponding peaks in the structure factor, $S(k)$, which can be experimentally probed using scattering techniques~\cite{araque2015modern}. We find that $g_{ZZ}(r)$ and $g_{NN}(r)$ individually capture the key structural features of $S(k)$: their oscillation periods ($l_o$) correspond well with the peak positions in $S(k)$, following the relation $k_o\approx2\pi/l_o$. $\pyrrtfsi$ is known to exhibit two characteristic peaks in $S(k)$~\cite{amith2020pictorial,umebayashi2011liquid}. The oscillation in $g_{ZZ}(r)$ is associated with the charge alternation, expecting the \textit{charge} peak located at $k_{peak}=2\pi/l_Z\approx8$~nm$^{-1}\approx k_{char}$, which is known to be associated with structural signatures of ionic liquids that are sensitive to the molecular characteristics of constituent ions~\cite{McDaniel2019}. Similarly, the oscillation of $g_{NN}(k)$ is related to the short-range ion packing, expecting the \textit{adjacency} peak in $S(k)$ at $k_{peak}\approx14$~nm$^{-1}\approx k_{adj}$. The so-called \textit{pre}-peak, associated with polarity alternation (typically $k_{pre}<6$~nm$^{-1}$), is absent, as $\pyrrtfsi$ contains a short alkyl chain that lacks sufficient spatial separation for long-range ordering. Further exploration of the scattering function is beyond the scope of the present study, but could be informative in understanding structural features of the IL-in-salt regime, particularly in a low-$k$ regime.

Figure~\ref{fig:gzzgnn}D displays the local charge atmosphere ($c^Q(r)$, Eq.~\ref{eq:cQr}) of the three constituent ions, all of which reflect the oscillatory exponential decays observed in $g_{ZZ}(r)$. At all $\xli$, the charge atmospheres exhibit clear oscillations whose amplitudes exceed the magnitude of the central ion’s charge (+1 or –1 e) and alternate between opposite signs, indicating the overcompensation similar to overscreening at an electrochemical interface~\cite{bazant2011double, Feng2019}. The amplitude of the oscillations decreases with increasing $\xli$, while the oscillation period remains nearly unchanged, consistent with the behavior observed in $g_{ZZ}(r)$. Interestingly, the oscillation period is found to be species-insensitive, in agreement with the assumptions of A. Lee et al.~\cite{Lee2017a}. Accordingly, we find that the ionic atmospheres can be well reconstructed using the $l_Z$ and $\lambda_Z$ parameters obtained from $g_{ZZ}(r)$ (see Table~\ref{si:tab:ionatm} in the SI). Only the phase of the oscillation is species-dependent, determining whether each species contributes constructively or destructively to the total scattering function~\cite{keblinski2000molecular,zeman2021ionic,Coupette2018,Krucker-Velasquez2021}. These findings highlight the collective nature of strong ion correlations in salt-in-ionic liquids. We again note that $g_{ZZ}(r)$ and $g_{NN}(r)$ were computed using the centers of charge and mass of the ions, respectively.

\begin {figure*}[t]
\centering
\includegraphics[width=6in]{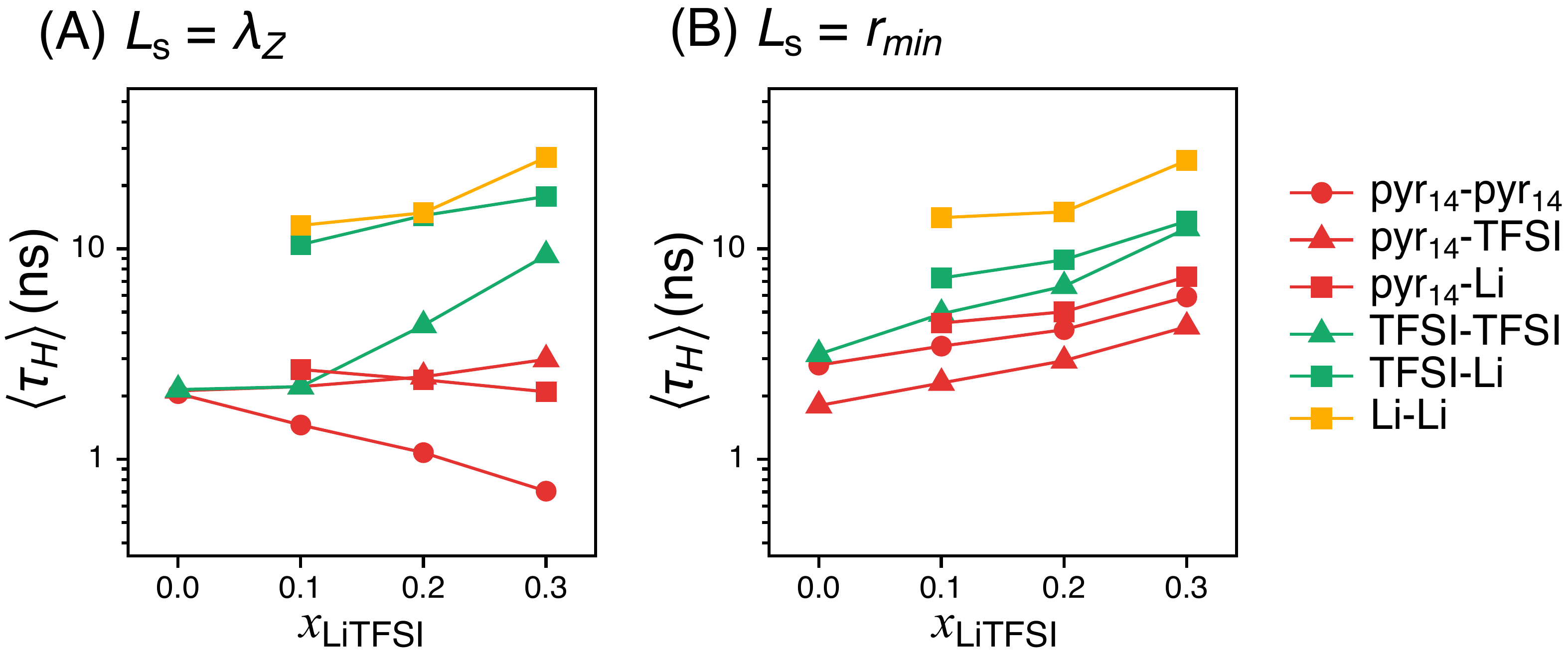}
\caption{Mean ion-pair lifetime ($\langle\tau_H\rangle$), estimated using $H(t)$ (Eq.~\ref{eq:pairdyn}), at various $\xli$ concentrations. Ion pairs are identified using a distance-based criterion $L_s$ in two ways: (A) $L_s=\lambda_Z$ and (B) $L_s=r_{\text{min}}$, the first minimum of each $g_{\alpha\beta}(r)$ (see Fig.~\ref{si:fig:rdf} in the SI). Different colors represent different ion pairs; in particular, red denotes those involving $\pyrr$.}
\label{fig:lifetime}
\end{figure*}

The observed decrease in $\lambda_Z$ with increasing $\xli$ aligns with predictions of ion cluster theory for sodium salt-doped ionic liquids~\cite{zhang2024long}, though the absolute values of $\lambda_Z$ remain an order of magnitude larger. The recently developed ion cluster theory~\cite{McEldrew2021,Goodwin2023,Kornyshev2022, mceldrew2020theory,zhang2024long} have been successful across several systems, including neat ionic liquids~\cite{Kornyshev2022}, superconcentrated electrolytes~\cite{mceldrew2020theory, McEldrew2021a}, and salt-in-ionic liquids~\cite{McEldrew2021b} by adopting the concept of gelation from polymer physics, and modeling ion associations as Cayley tree-like clusters. The theory proposed that the multivalent coordination of alkali ions with molecular anions strongly promotes ion cluster formation, effectively increasing the ionic strength. Specifically, negatively charged \ce{Na-TFSI} clusters were considered in the theory, as the association between \ce{TFSI} and molecular cations is weak, thereby increasing the ionic strength and reducing $\lambda_Z$. In this work, we also observe the formation of ion clusters, particularly those involving $\li$ and $\tfsi$ (see Fig.~\ref{si:fig:cluster_lian} in the SI); these Li-TFSI clusters are asymmetric and carry a net negative charge, consistent with previous studies~\cite{McEldrew2021b,zhang2024long,Molinari2019}, contributing to the so-called negative lithium transference number, as will be discussed in Section~\ref{sec:diff}. Despite the prevalent small clusters, we do not observe any \textit{percolating}  Li-TFSI clusters that span the simulation box, suggesting that the LILs over the investigated $\xli$ range remain in the pre-\textit{gel} regime. We should note that SFB experiments~\cite{zhang2024long} reported an \textit{increase} in the decay length of surface forces with increasing \ce{NaTFSI} concentration, a trend not captured by our bulk simulations or the ion cluster theory.

\subsection{$x_{\litfsi}$-dependent $\lambda_Z$ disentangles correlations in ion motions}\label{sec:diff}

Figure~\ref{fig:cond}A displays the computed ionic conductivity ($\sigma$) and its ratio to the value predicted by the Nernst–Einstein approximation ($\sigma_{NE}$). The $\sigma$ and $\sigma_{NE}$ were computed according to Eq.~\ref{eq:gk_conductivity}, and Eq.~\ref{eq:ne_conductivity}, respectively. As $\xli$ increases, both $\sigma$ and $\sigma_{NE}$ decrease (see Table~\ref{si:tab:props} in the SI), which can be attributed to the higher density and viscosity (reduced ion self-diffusion) and the increased dynamic correlations~\cite{Kubisiak2021,Molinari2019,gouverneur2018negative,Wettstein2022,zhang2024long}. At $\xli = 0.3$, $\sigma$ is reduced to approximately half of its value in pure \pyrrtfsi. Previous experiments~\cite{pitawala2012phase} reported a similar decreasing trend in $\sigma$ up to $\xli\approx0.4$ in salt-in-ionic liquid systems. Despite the increasing glass transition temperature ($T_g$), the $T_g$-corrected conductivity further suggested that these systems share similar fragility.

Figure~\ref{fig:cond}A displays the calculated \textit{ionicity} ($\sigma/\sigma_{NE}$) as a function of $\xli$. Typically, $\sigma/\sigma_{NE}<1$ due to the correlated ion motions that hinder ion transport; $\sigma/\sigma_{NE}=1$ when the correlations in ion motion have no contribution. The pure \pyrrtfsi\textcolor{red}{~}exhibits $\sigma/\sigma_{NE} \approx 0.75$, consistent with previous studies~\cite{Haskins2014}. At $\xli = 0.1$, $\sigma/\sigma_{NE}$ decreases due to enhanced ionic correlations (\textit{e.g.}, small Li-TFSI clusters). However, further addition of $\litfsi$ increases $\sigma/\sigma_{NE}$ to approximately 0.8 in the range $0.2 \leq \xli \leq 0.3$, suggesting a reduced contribution of correlated ion motions.
This trend of ionicity with $\xli$ appears non-monotonic, although the statistical uncertainty limits a definitive conclusion. We posit that the seemingly non-monotonic behavior arises from the liberation of \ce{[pyr14]+} ions due to strong correlations between \ce{Li+} and TFSI$^-$ ions, as discussed below. We note that such limited statistics in ionicity are not uncommon in ionic liquid systems~\cite{Lesch2014}, where dynamic correlations between different ionic species contribute significantly to ionic conductivity, and thus to ionicity.

Figure~\ref{fig:cond}B shows the computed transference number ($t_{\alpha}$) of all three ions, which indicates the fraction of the contribution of each ion to $\sigma$. $\Delta t_{\alpha} = t_{\alpha}-t_{\alpha}(\xli=0)$ is the change in transference number of $\alpha$ species relative to the value of neat $\pyrrtfsi$. By definition, $t_{\ce{Li}} = 0$ for pure \pyrrtfsi. In all compositions studied, $t_{\ce{Li}}$ is negative, which has been widely observed in salt-in-ionic liquid systems~\cite{gouverneur2018negative,Molinari2019,molinari2020chelation}. This is associated with the formation of asymmetric Li-anion clusters (see Fig.~\ref{si:fig:cluster_lian} in the SI); such clusters (\textit{e.g.}, [TFSI–Li–TFSI]$^{-1}$) carries a net negative charge, causing $\li$ ions to migrate toward the "wrong" electrode.

At $\xli = 0.3$, there appears a turnover in $t_{\ce{Li}}$, which is closely related to the effective charge transferred ($q_{\text{eff}}$)~\cite{Molinari2019,McEldrew2021b}, which is negative in salt-in-ionic liquid systems but gradually increases with increasing $\xli$, eventually becoming positive in ionic liquid-in-salt systems. Our calculation of $q_{\text{eff}}$ (see the SI for further details) follows the similar trend of $t_{\ce{Li}}$ across all $\xli$. At high $\xli$, larger Li–TFSI clusters form that barely contribute to $\sigma$ due to their extremely slow diffusion; only free ions and small ion clusters contribute to $\sigma$. In contrast, small and asymmetric Li–TFSI clusters contribute negatively to $\sigma$ at low $\xli$. 

With varying $\xli$, the transference numbers of $\pyrr$ and $\tfsi$ should also change, maintaining the relation $\sum_{\alpha} t_{\alpha}=1$ (Fig.~\ref{fig:cond}B). Both $t_{\ce{pyr14}}$ and $t_{\ce{TFSI}}$ increase with increasing $\xli$, consistent with the negative $t_{\ce{Li}}$, exhibiting a similar non-monotonic trend. This behavior can be rationalized by the strong $\ce{Li–TFSI}$ associations: The formation of $\ce{Li–TFSI}$ clusters effectively liberates $\pyrr$ ions, increasing $t_{\ce{pyr14}}$ relative to the neat $\pyrrtfsi$. Despite the strong interactions of $\tfsi$ with $\li$, $t_{\ce{TFSI}}$ also increases due to the multivalent coordination of $\li$ in forming ion clusters. The turnover in the transference number is attributed to the aggregation of these small $\ce{Li–TFSI}$ clusters, even though no percolating ion network is observed in this work. This trend in transference numbers further reflects the complex nanoscopic organization of LILs and provides a connection between ionic clusters and ion transport.

\begin {figure*}[th]
\centering
\includegraphics[width=7in]{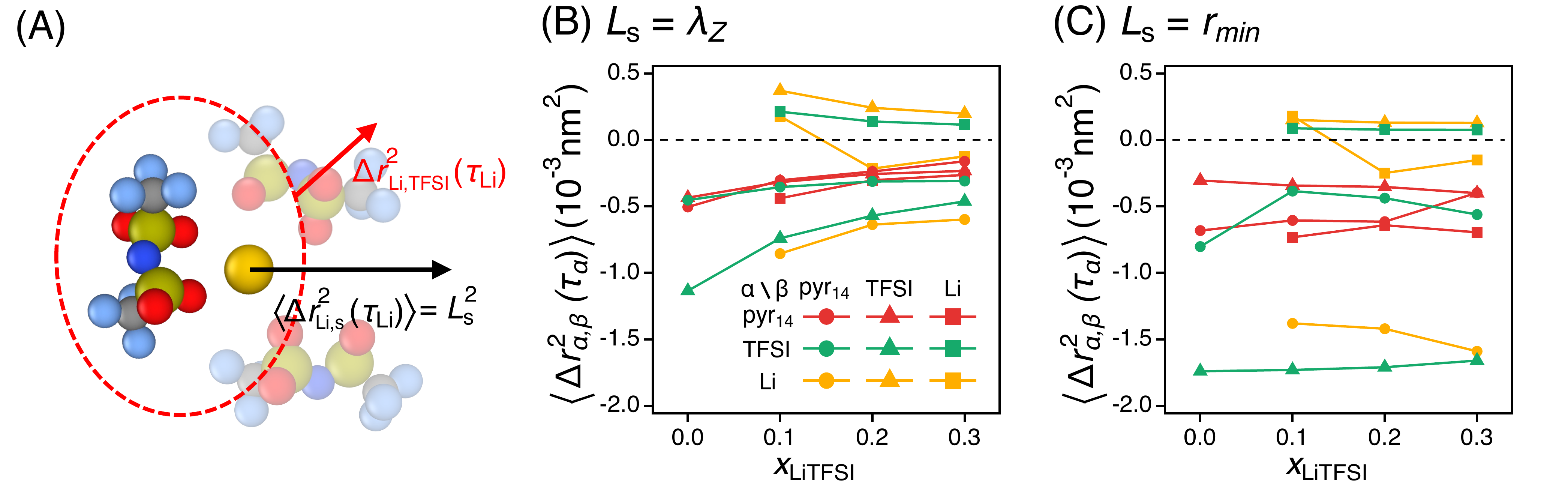}
\caption{Correlated ion motion relative to self-diffusion at various $\litfsi$ doping fractions, $\xli$. (A) Schematic illustration of $\li$ diffusion with coordinating $\tfsi$ ions. $\tau_{\alpha}$ denotes the time at which the self mean-squared displacement (MSD) of species $\alpha$ reaches the square of a distance criterion, $L_s$. (B, C) Cross MSDs, $\langle\Delta r^2_{\alpha,\beta}(\tau_\alpha)\rangle$, between species $\alpha$ and $\beta$, where $\alpha$ is the species of interest and $\beta$ exhibits correlated diffusion with $\alpha$. The distance criterion is defined as (B) $L_s=\lambda_Z$ and (C) $L_s=r_{\text{min}}$. See the main text for further details. The values of $\lambda_Z$ and $r_{min}$ are listed in Fig.~\ref{fig:gzzgnn}C and Table~\ref{si:tab:firstshell}, respectively.}
\label{fig:correlative_transport}
\end{figure*}

The structural organization of charge and its dynamics are intimately related~\cite{lundin2022ion,eliasen2024origin,ghorai2020equilibrium}. For example, simulations of imidazolium-based ILs~\cite{ghorai2020equilibrium}, based on analysis of the dynamic charge correlation structure factor, identified two distinct dynamical regimes in charge-density fluctuations: a faster, stretched-exponential decay and a slower, exponential relaxation associated with charge-alternating structures. Notably, the peak corresponding to charge oscillations in the charge-charge structure factor, which is the Fourier transform of $g_{ZZ}(r)$, exhibited the slowest relaxation. This behavior is consistent with de Gennes narrowing, demonstrating the intrinsic relationship between charge structure and dynamics in ILs.

In this work, to further elucidate the connection between electrostatic screening and ion transport, we estimate how long ion pairs persist using the time correlation function $H_{\alpha\beta}(t)$ (Eq.~\ref{eq:pairdyn}, see Fig.~\ref{si:fig:pairdyn_lz} in the SI). We determine whether two ions form a pair based on two different distance-based criteria: $\lambda_Z$ and the first minimum($r_{\text{min}}$) of the ion-ion RDF. Our choice of $\lambda_Z$ follows at least two reasons: (i) $\lambda_Z$ measures the distance over which charge-charge correlations persist, and (ii) $\lambda_Z$ is found to be species-insensitive (Fig.~\ref{fig:gzzgnn}C), implying that it can be consistently applied to all constituent ions. In contrast, $r_{\text{min}}$ is the one commonly used in previous studies~\cite{Self2019,Zhang2015}. Fig.~\ref{fig:lifetime} displays the mean ion-pair lifetime ($\langle\tau_H\rangle$), computed using $H_{\alpha\beta}(t)$ with each of the distance-based criteria. It is evident that the observed pair dynamics, spanning more than an order of magnitude, are sensitive to the choice of the distance criterion. Fig.~\ref{fig:lifetime}A clearly shows the pair-dependent results of $\langle\tau_H\rangle$ as a function of $\xli$. The decay of $H_{\alpha\beta}(t)$ is also pair-dependent (see Fig.~\ref{si:fig:pairdyn_lz} in the SI). Consistent with the formation of Li–TFSI clusters (Fig.~\ref{si:fig:cluster_lian} in the SI), the Li–TFSI, Li–Li, and TFSI–TFSI pairs persist longer as $\xli$ increases. Among all pairs, the Li–Li correlation time is the slowest across all $\xli$, consistent with the fact that $\li$ shows the slowest dynamics compared to the other molecular ions (see Fig.~\ref{si:fig:msd}). At $\xli=0.3$, $\langle\tau_H\rangle$ of Li–TFSI becomes comparable to that of Li–Li, being on the order of 10 ns (see Table~\ref{si:tab:hfit_lz}). This result, again, support the correlated diffusion of $\li$-$\tfsi$ clusters contribute significantly to ionic conductivity according to the ion cluster theory~\cite{McEldrew2021b}. 

Interestingly, $\langle\tau_H\rangle$ of $\ce{pyr_{14}-pyr_{14}}$ and $\ce{Li-pyr_{14}}$ pairs , when defined using the screening length $\lambda_Z$, \textit{decrease} with increasing $\xli$ (Fig.~\ref{fig:lifetime}A). This implies that the ion clusters involving $\pyrr$ contribute less to the total ionic conductivity with increasing $\xli$ according to the ion cluster theory, which posits that only clusters with lifetimes significantly exceeding their velocity correlation times contribute to ion conduction. A similar decrease in dynamic correlations of $\pyrr$ was reported by previous simulation studies~\cite{shankar2024impact,Wettstein2022,McEldrew2021b}, inferred indirectly from transport signatures. These shorter lifetimes further indicate the liberation of $\pyrr$ cations, reflecting a shift in balance toward strengthened correlations between Li$^+$ and TFSI$^-$ ions. This counterintuitive dynamic behavior of $\pyrr$ is likely a key driver of the increasing $\sigma_{\mathrm{NE}}/\sigma$ with $\xli$ (Fig.~\ref{fig:cond}B). 

In the case of $r_{\text{min}}$ (Fig.~\ref{fig:lifetime}B), however, all ion pairs persist longer at higher $\xli$, reflecting the increased overall viscosity with more $\litfsi$ salt. We note that $r_{\text{min}}$ is largely insensitive to $\xli$ in contrast to $\lambda_Z$. Such a criterion results in slower decays of $H_{\alpha\beta}(t)$ across all ion pairs (see Fig.~\ref{si:fig:pairdyn_rmin} in the SI), increasing $\langle\tau_H\rangle$ with increasing $\xli$. The relative ordering of $\langle\tau_H\rangle$ across all ion pairs remains largely unchanged with $\xli$, \emph{i.e.}, the species-specific contributions captured with $r_{\text{min}}$ are hardly dependent on $\xli$. We also obtain similar results with another $\xli$-insensitive species-dependent criterion, the $\lambda_Z$ at $\xli=0$ (see Fig.~\ref{si:fig:pairdyn_fixl} in the SI). These results in Fig.~\ref{fig:lifetime} highlight the central role of $\lambda_Z$ in capturing species-specific dynamic correlations and disentangling their contributions to the total ionic conductivity in complex ionic liquid systems. We note that while the decay of $H_{\alpha\beta}(t)$ is related to the disintegration dynamics of an ion solvation shell, the two are not identical~\cite{Wettstein2022}.

We further investigate the role of $\lambda_Z$ as a length scale to disentangle species-dependent contributions to the ionic conductivity ($\sigma$). Correlations in ion motions, along with reduced ion self-diffusion, contribute to the decrease ionic conductivity as reflected in $\sigma/\sigma_{NE}$ (Fig.~\ref{fig:cond}). The cross-correlations ($\langle\Delta r^2_{\alpha,\beta}(t)\rangle$) in ion motions depend on several factors, such as $\litfsi$ concentration and ion species, and vary across different length and time scales. As illustrated in Fig.~\ref{fig:correlative_transport}A, we measure $\langle\Delta r^2_{\alpha,\beta}(t)\rangle$ up to a point where the $\alpha$ ion, on average, diffuses a certain distance, $L_s$ ($L_s=\lambda_Z$ in the case). Here, $\alpha$ and $\beta$ represent ion species. Accordingly, the characteristic time scale ($\tau_\alpha$) is defined based on self-diffusion of $\alpha$ species as $\langle\Delta r^2_{\alpha,s}(\tau_\alpha)\rangle = L_s^2$. We note that $\langle\Delta r^2_{\alpha,\beta}(\tau_\alpha)\rangle \neq \langle\Delta r^2_{\beta,\alpha}(\tau_\beta)\rangle$, since $\tau_\alpha \neq \tau_\beta$ in general. We again adopt two length scales for $L_s$: (i) $\lambda_Z$ (Fig.~\ref{fig:correlative_transport}B) and (ii) $r_{min}$ (Fig.~\ref{fig:correlative_transport}C) to provide a framework for connecting electrostatic screening and ion transport in various ionic liquid systems. We note that the results of $\langle\Delta r^2_{\alpha,\beta}(\tau_\alpha)\rangle$ for all $\alpha$-$\beta$ ion pairs shown in Figs.~\ref{fig:correlative_transport} are not distance-resolved, but rather represent the average contributions from all $\alpha$-$\beta$ pairs.

Figure~\ref{fig:correlative_transport}B summarizes the results of $\langle\Delta r^2_{\alpha,\beta}(\tau_\alpha)\rangle$ for all $\alpha$-$\beta$ ion pairs with the distance scale of $\lambda_Z$. It is clear that the magnitude and sign of $\langle\Delta r^2_{\alpha,\beta}(\tau_\alpha)\rangle$ depend on the choice of the $\alpha$-$\beta$ pair and the $\litfsi$ concentration. $\lambda_Z$ decreases with increasing $\xli$. Among all the ion pairs, the $\li$-$\tfsi$ pairs diffuse together (positive $\langle\Delta r^2_{\text{Li,TFSI}}(\tau_{\text{Li}})\rangle$ and $\langle\Delta r^2_{\text{TFSI,Li}}(\tau_{\text{TFSI}})\rangle$), while all other pairs exhibit anti-correlated motion, consistent with previous studies~\cite{Kubisiak2021,Wettstein2022}. The formation of Li-TFSI clusters strongly affects their correlated dynamics, as expected. We note that the negative $\langle\Delta r^2_{\ce{TFSI},\ce{TFSI}}(\tau_\alpha)\rangle$ indicates the contribution to a decrease in the ion conductivity. Notably, the absolute value of $\langle\Delta r^2_{\alpha,\beta}(\tau_\alpha)\rangle$ decreases with increasing $\xli$, regardless of the choice of the $\alpha$-$\beta$ pair. These reduced correlations in ion motion with decreasing $\lambda_Z$ are further consistent with the results in Wettstein et al.~\cite{Wettstein2022}, where lithium coupling factor was found to decrease with increasing $\xli$. 

In Fig.~\ref{fig:correlative_transport}C, instead of $\lambda_Z$, we adopt $L_s=r_{min}$ and correspondingly a conventional timescale $\tau_\alpha$, defined as the time over which the $\alpha$ ion, on average, diffuses $r_{\text{min}}$ of the $\alpha$-$\beta$ RDF, where $\beta$ ions solvate the $\alpha$ ion. $r_{\text{min}}$ is sensitive to the choice of a $\alpha$-$\beta$ pair, while remaining largely insensitive to $\xli$. It is again evident that $\langle\Delta r^2_{\alpha,\beta}(\tau_\alpha)\rangle$ depends strongly on the selected ion pair. However, at the length scale of $r_\text{min}$, the dependence of correlated motions on $\xli$ is not straightforward; most of $\langle\Delta r^2_{\alpha,\beta}(\tau_\alpha)\rangle$ is largely unchanged with $\xli$. For instance, Li-TFSI (yellow triangles and green squares) and TFSI-TFSI (green triangles) do not depends on $\xli$ in spite of the fact that the contributions of these pairs to the ionic conductivity decrease with $\xli$ (Fig.~\ref{si:fig:contribution} in the SI). This further emphasizes that our analysis, based on the $\xli$-dependent $\lambda_Z$, enables a quantitative comparison of the motion of correlated ions across LILs with variations $\xli$, while minimizing the confounding effects of changes in the overall viscosity.

\section*{Conclusions and Future Directions}\label{sec:conclusions}
In this work, we performed atomistic molecular dynamics simulations of $\pyrrtfsi$ doped with $\litfsi$ at molar fractions $\xli = 0, 0.1, 0.2$, and 0.3 to investigate the connection between electrostatic screening and collective ion transport. In addition to the changes in short-range ion solvation structures, we find that both $g_{ZZ}(r)$ and $g_{NN}(r)$ exhibit the characteristics of the oscillatory exponential decay, indicating that LILs are charge- and mass-dense systems. The oscillation of each correlation function, insensitive to $\xli$, effectively captures the nanostructures of the LILs, associated with charge alternation and spatial adjacency, respectively. However, the decay lengths of the correlation functions show distinct trends upon $\litfsi$ doping: $\lambda_Z$ decreases with increasing $\litfsi$ concentration, whereas $\lambda_N$ remains nearly unchanged. The decreased $\lambda_Z$ with increasing $\xli$ is consistent with the ion cluster theory for sodium salt-doped ionic liquids~\cite{zhang2024long}, which predicts the formation of ion clusters influencing transport properties. Due to the strong interactions between $\li$ and $\tfsi$, ionic clusters form that are asymmetric, \emph{e.g.}, a triplet of [$\tfsi$-$\li$-$\tfsi$]$^{-1}$, which carry a net negative charge. Within the studied $\xli$ range, the ion clusters remain small with no percolating ionic network observed.

Regarding collective ion transport, we find that the $\xli$-sensitive $\lambda_Z$ serves as a central length scale for disentangling species-specific contributions to collective ion transport upon $\litfsi$ doping. We utilize $\lambda_Z$ as a distance-based criterion for determining ion pairs. Due to the enhanced propensity to form $\ce{Li-TFSI}$ clusters, the lifetime of ion pairs involving $\li$ and $\tfsi$ increases, while ion pairs involving $\ce{pyr_{14}}$ decay faster with increasing $\litfsi$ concentration. The results are found sensitive to the choice of the distance criterion. A $\xli$-insensitive distance criteria, such as the first minimum of ion-ion RDFs, and $\lambda_Z$ of pure $\pyrrtfsi$, lead to species-\textit{independent} slowing down of ion pair dynamics with increasing $\litfsi$ concentration. Our further analysis suggests that the stronger $\ce{Li-TFSI}$ correlations at higher $\xli$ concentrations promote the formation of ionic clusters, liberating $\ce{pyr_{14}}$ cations, and thereby enhancing the overall ionicity of the LILs over the investigated $\xli$ range. Therefore, the approach in this study provides a framework for connecting electrostatic screening and ion transport in LILs. Applying this framework across a broader range of ionic liquid systems could further elucidate the connection between screening length and ion transport in future studies.

One direction for future research is to investigate the universality of underscreening behavior in alkali salt-doped ionic liquids (ILs), particularly in the concentrated regime and across different salt chemistries. The nature of electrostatic screening in ILs remains puzzling, especially the experimentally observed universal cubic scaling of underscreening in SFB experiments~\cite{Gebbie2015,Jager2023,Lee2017a}. While most simulations~\cite{coles2020correlation,zeman2021ionic,Krucker-Velasquez2021} and classical DFT studies~\cite{Adar2019,Cats2021,Rotenberg2018} reproduce underscreening trends, they typically underestimate the scaling exponent, reporting values between 1 and 2 rather than the experimentally observed value of 3. Recent studies~\cite{hartel2023anomalous,Kumar2022,wang2024structure,elliott2024known} have suggested that interface and confinement effects, such as enhanced ion association in the narrow gaps between surfaces in SFB experiments, may account for this discrepancy.

Another potential direction for future work involves investigating how subdiffusive signatures in short-time MSDs influence long-time ionic transport. We find that the associated timescale relevant to $\lambda_Z$, as indicated by both self and cross parts of the MSDs, lies in a subdiffusive regime, even though ionic conductivity is obtained in the diffusive regime. For instance, at short times ($t \lesssim 1$ ns), the collective MSD, similar to the lithium self-diffusion MSD, exhibits subdiffusive behavior: $\langle\Delta r^2_{\text{Li,TFSI}}(t)\rangle \sim t^{\beta}$ with $0.5\lesssim\beta<1$. Such a relevance of subdiffusion to long-time ion conductivity was also suggested in the context of flow-like ion transport~\cite{Wettstein2022,diddens2023hydrodynamic}. Thus, future work could explore how short-time subdiffusive dynamics connects to long-time ionic transport, particularly in relation to structural features.

\section*{Supplementary Material}
The supplementary material provides more details on the computed properties, fit results, RDF between ions, the charge atmosphere derivation using the Debye-Huckel approximation, lithium-ion solvation environment, Li-TFSI ion clusters analysis, ion transport results, and ion pair dynamics.

\begin{acknowledgments}
The authors thank Chung Bin Park for his feedback on the initial version of the manuscript. This work was supported by the National Research Foundation of Korea (NRF) grant funded by the Korea government (MSIT) (RS-2024-00338551).
\end{acknowledgments}

\section*{Data Availability Statement}
The data that support the findings of this study are available from the corresponding author upon reasonable request.

%\bibliography{main}


\end{document}

\preprint{%AIP/123-QED
}

\title[]{
Supporting Information for \\ Bridging Electrostatic Screening and Ion Transport in Lithium Salt-Doped Ionic Liquids}
\author{Hyungshick Park$^{a}$, Bong June Sung$^{a,b,\dagger}$, and Jeongmin Kim$^{c,\ast}$}
\affiliation{$^a$Department of Chemistry, Sogang University, Seoul 04107, Republic of Korea \\
 $^b$Institute of Biological Interfaces, Sogang University, Seoul 04107, Republic of Korea\\
 $^c$Department of Chemistry Education and Graduate Department of Chemical Materials, Pusan National University, Busan 46241, Republic of Korea}
\email{jeongmin@pusan.ac.kr, $^\dagger$bjsung@sogang.ac.kr}

%\author{Hyungshick Park}
%\affiliation{Department of Chemistry, Sogang University, Seoul 04107, Republic of Korea}

%\author{Bong June Sung$^{\dagger}$}
%\affiliation{Department of Chemistry, Sogang University, Seoul 04107, Republic of Korea}
%\affiliation{Institute of Biological Interfaces, Sogang University, Seoul 04107, Republic of Korea}
%\email{bjsung@sogang.ac.kr}

%\author{Jeongmin Kim$^{\ast}$}
%\affiliation{Department of Chemistry Education and Graduate Department of Chemical Materials, Pusan National University, Busan 46241, Republic of Korea}
%\email{jeongmin@pusan.ac.kr}

\maketitle

\newpage

\renewcommand{\thesection}{S\arabic{section}}
\setcounter{section}{0}
\renewcommand{\thefigure}{S\arabic{figure}}
\setcounter{figure}{0}
\renewcommand{\thetable}{S\arabic{table}}
\setcounter{table}{0}
\renewcommand{\theequation}{S\arabic{equation}}
\setcounter{equation}{0}

\renewcommand{\thesection}{S\arabic{section}}
\setcounter{section}{0}
\renewcommand{\thesubsection}{S\arabic{subsection}}
\setcounter{subsection}{0}
\renewcommand{\thefigure}{S\arabic{figure}}
\setcounter{figure}{0}
\renewcommand{\thetable}{S\arabic{table}}
\setcounter{table}{0}
\renewcommand{\theequation}{S\arabic{equation}}
\setcounter{equation}{0}

\section{Computed properties and fit results}

\begin{table*}[h]
%\centering
\begin{ruledtabular}
\begin{tabular}{r|rrrrrrrrr}
$x_{\litfsi}$ & $\rho_{\text{ion}}$ & $c_{\text{ion}}$ & $D_{\ce{pyr_{14}}}$ & $D_{\ce{TFSI}}$ & $D_{\ce{Li}}$ & $\sigma_{NE}$ & $\sigma$ & $t_{Li}$ & $t^{NE}_{Li}$ \\
 - &  (g/cm$^3$) & (M) & (nm$^2$/ns) & (nm$^2$/ns) & (nm$^2$/ns) & (mS/cm) & (mS/cm) & - & - \\ \hline
0.0      & 1.27 & 3.0  & 0.10 & 0.10 & -    & 17.3  & 13.1  & - & - \\
0.1      & 1.30 & 3.2  & 0.08 & 0.07 & 0.04 & 13.5  & 8.9  & -0.02 & 0.029 \\
0.2      & 1.33 & 3.3  & 0.06 & 0.05 & 0.03 & 9.8  & 8.5  & -0.07 & 0.054 \\
0.3      & 1.37 & 3.6  & 0.04 & 0.03 & 0.02 & 6.5  & 5.5  & -0.04 & 0.085 \\ 
\end{tabular}
\caption{Summary of computed properties for pure and lithium-doped \pyrrtfsi~systems investigated in this work. Here, $x_{\litfsi}$, $\rho_{\text{ion}}$, and $c_{\text{ion}}$ represent the molar fraction of \litfsi, and the mass density and number density of total ions, respectively. See the main text for further details}
\label{si:tab:props}
\end{ruledtabular}
\end{table*}

\begin{table}[h]
%\begin{center}
\begin{ruledtabular}
\begin{tabular}{rrrrr}
%\hline
$x_{\litfsi}$ & $\lambda_Z$ & $l_Z$ & $A_Z$ & $\phi_Z$ \\ \hline
0.0     & 1.17  & 0.79 &  0.98 &  1.89\\
0.1     & 0.94  & 0.80 &  0.88 &  2.80\\
0.2     & 0.83  & 0.80 &  0.63 &  2.12\\
0.3     & 0.75  & 0.79 &  0.43 &  1.95\\ \hline\hline
$x_{\litfsi}$ & $\lambda_N$ & $l_N$ & $A_N$ & $\phi_N$ \\ \hline
0.0     & 0.45  & 0.45 &  1.3 &  1.32 \\
0.1     & 0.43  & 0.47 &  1.2 &  2.20 \\
0.2     & 0.43  & 0.48 &  0.9 &  2.60 \\
0.3     & 0.44  & 0.48 &  0.7 &  2.76 \\ %\hline
\end{tabular}
\caption{Fit results for $\ln(|g_{ZZ}(r)\cdot r|)$ and $\ln(|(g_{NN}(r)-1)\cdot r|)$ at various $\xli$, obtained using Eq.~\ref{eq:gzz_fit} and Eq.~\ref{eq:gnn_fit} (Fig.~\ref{fig:gzzgnn}A and B in the main text). The fitting parameters $\lambda$, $l$, $A$, and $\phi$ correspond to the decay length, oscillation period, amplitude, and phase, respectively. Subscripts $Z$ and $N$ denote results for $\ln(|g_{ZZ}(r)\cdot r|)$ and $\ln(|(g_{NN}(r)-1)\cdot r|)$, respectively. Fits were performed over the ranges $r \in (1, 3)$ nm for charge–charge and $r \in (1, 2)$ nm for density–density correlations.}
\label{si:tab:gzzgnn}
%\end{center}
\end{ruledtabular}
\end{table}

\begin{table*}[!htbp]
\begin{ruledtabular}
\begin{tabular}{r|rrr|rrr|rrr}

         & \multicolumn{3}{r|}{\li} & \multicolumn{3}{r|}{\pyrr} & \multicolumn{3}{r}{\tfsi} \\ \hline
$x_{\litfsi}$ & range (nm) & $A_Q$ & $\phi_Q$ & range (nm) & $A_Q$ & $\phi_Q$ & range (nm) & $A_Q$ & $\phi_Q$ \\ \hline
0.0      &     -      &   -    &    -     & (0.9,3.7) & -0.001 & 1.752  & (0.9,3.7) &  0.001 & 1.759  \\
0.1      & (0.9,3.7) & 0.003 &  6.638  & (0.9,3.7) & -0.003 & 1.928  & (0.9,3.7) & -0.003 & 1.928  \\
0.2      & (1.4,3.7) &  0.004 &  6.539  & (2.0,3.8) & -0.006 & 1.749  & (2.0,3.8) & 0.003 & 1.865  \\
0.3      & (2.3,3.7) & -0.005 & 15.700  & (2.4,3.8) & -0.006 & 1.759  & (2.4,3.8)& 0.003 & 1.546  \\ 
\end{tabular}
\caption{Fit results for the ion atmospheres of $\li$, $\pyrr$, and $\tfsi$ at various $\xli$, using Eq.~\ref{eq:cq_fit} (Fig.~\ref{fig:gzzgnn}C in the main text). The decay length $\lambda_Z$ and oscillation period $l_Z$ are fixed to the values listed in Table~\ref{si:tab:gzzgnn}, while the amplitude $A_Q$ and phase $\phi_Q$ are fitted parameters. Fits were performed over the indicated range.}
\label{si:tab:ionatm}
\end{ruledtabular}
\end{table*}

\clearpage

\section{Radial distribution functions between the ions}

% Structure changes of ions
\begin {figure*}[htb]
\centering
\includegraphics[width=5.in]{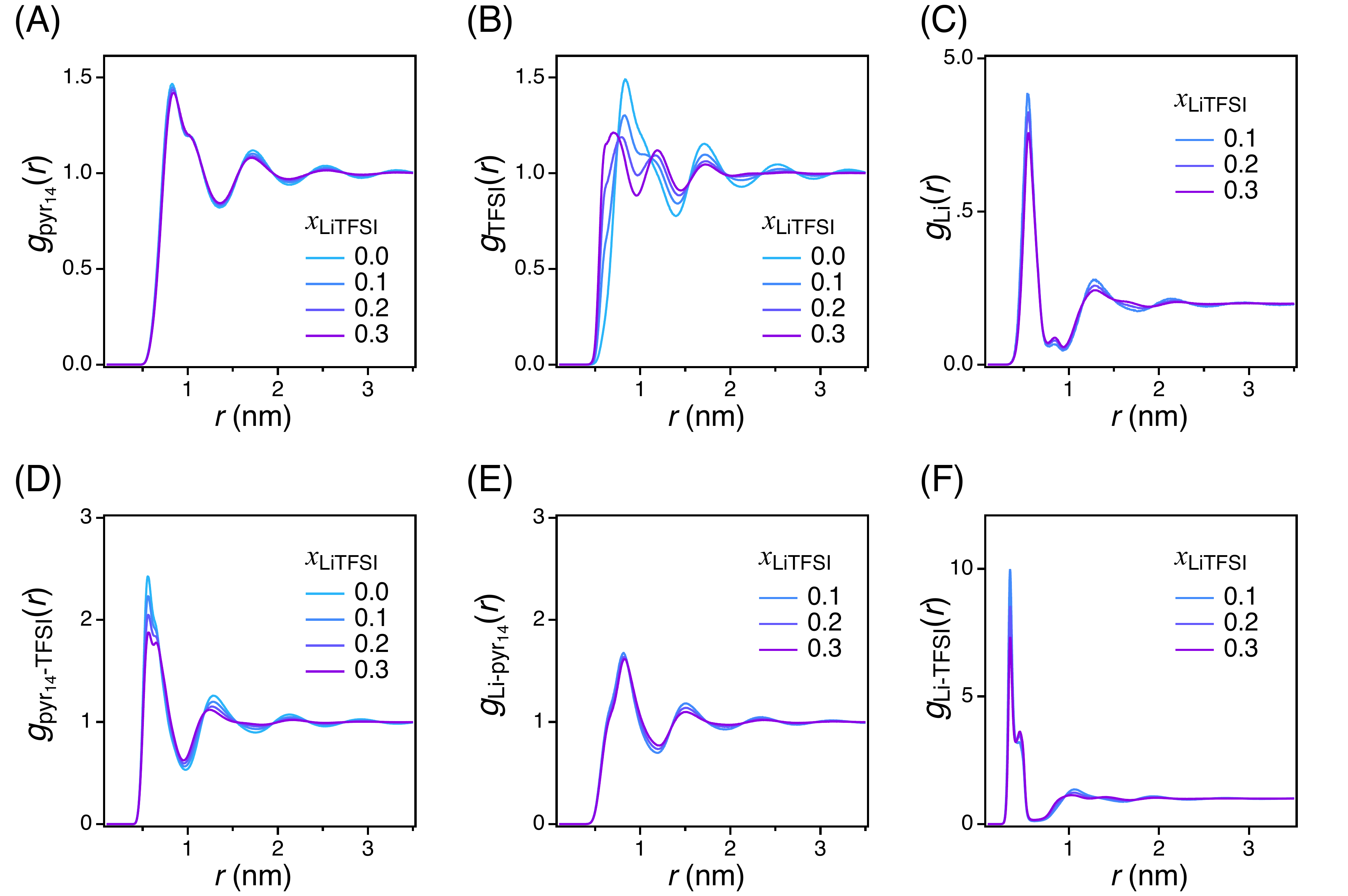}
\caption{
Radial distribution functions (RDFs, $g(r)$) between the ions at various $\xli$. The upper panels present $g(r)$ between the same ion species: (A) $\pyrr$, (B) $\tfsi$, and (C) $\li$. The lower panels present $g(r)$ between the different ion species: (D) \ce{pyr14}-\ce{TFSI}, (E) $\ce{Li}$-$\ce{pyr14}$, and (F) $\ce{Li}$-$\ce{TFSI}$. The RDFs were computed using the centers of mass of the ions.
}
\label{si:fig:rdf}
\end{figure*}

\begin{table}[!htbp]
\begin{ruledtabular}
\begin{tabular}{rr}

ion pair        & $r_{min}$ (nm) \\ \hline
$\pyrr$-$\pyrr$ & 1.36 \\
$\tfsi$-$\tfsi$ & 1.0 / 1.45 \\
$\li$-$\li$     & 0.95 \\
$\pyrr$-$\tfsi$ & 0.98 \\
$\li$-$\pyrr$   & 1.21 \\
$\li$-$\tfsi$   & 0.62 \\

\end{tabular}
\caption{
The size of the first solvation shell, $r_\text{min}$. All results are obtained using $g(r)$ in Fig.~\ref{si:fig:rdf}: $r_\text{min}$ is the first minimum of $g(r)$. In the case of $\ce{TFSI}$-$\ce{TFSI}$ pairs, we present distances of the first and the second shell due to the formation of $\ce{TFSI}-\ce{Li}-\ce{TFSI}$ triplets. The values of $r_\text{min}$ are used as a distance-based criterion for the calculation of the ion pair dynamics ($H(t)$)
}
\label{si:tab:firstshell}
\end{ruledtabular}
\end{table}

\newpage

\section{Derivation of $c^Q(r)$ according to the Debye-H\"uckel theory} \label{si:sec:cQr_DH}

According to the Debye--Hückel approximation, the radial distribution function is:
\begin{equation}
g_{\alpha\beta}(r) \approx 1 - \frac{Z_\alpha Z_\beta}{e^2}\frac{l_B}{r} \exp\left(-\frac{r}{\lambda_D}\right),
\end{equation}
where $l_B$ is the Bjerrum length and $\lambda_D$ is the Debye length. Then, the charge-charge correlation function is given as follows:
\begin{equation}
g_{ZZ}(r) \approx \frac{(Z_\alpha Z_\beta)^2}{e^2}\frac{l_B}{r} \exp\left(-\frac{r}{\lambda_D}\right).
\end{equation}
Substituting this into the definition of $c_{\alpha\beta}(r)$:
\begin{align}\label{eq:cDH}
c_{\alpha\beta}(r) 
&= \rho_\beta \int_0^r 4\pi r'^2 \left[1 - \frac{Z_\alpha Z_\beta}{e^2}\frac{l_B}{r'} e^{-r'/\lambda_D} \right] dr' \\
&= \rho_\beta \cdot \frac{4\pi}{3} r^3 
- 4\pi \frac{Z_\alpha Z_\beta}{e^2}\rho_\beta l_B \lambda_D^2 \left[ 1 - e^{-r/\lambda_D} \left( 1 + \frac{r}{\lambda_D} \right) \right]
\end{align}
Inserting this into the expression for $c^Q_\alpha(r)$:
\begin{align}
c^Q_\alpha(r) 
&= Z_\alpha + \sum_\beta Z_\beta\, c_{\alpha\beta}(r) \\
&= Z_\alpha - \sum_\beta Z_\beta \cdot 4\pi \rho_\beta l_B \frac{Z_\alpha Z_\beta}{e^2} \lambda_D^2 \left[1 - e^{-r/\lambda_D} \left(1 + \frac{r}{\lambda_D} \right) \right]  \\
&= Z_\alpha - Z_\alpha \cdot \frac{4\pi l_B \lambda_D^2}{e^2} \sum_\beta \rho_\beta Z_\beta^2 \left[1 - e^{-r/\lambda_D} \left(1 + \frac{r}{\lambda_D} \right) \right] \\
&= Z_\alpha - Z_\alpha \left[1 - e^{-r/\lambda_D} \left(1 + \frac{r}{\lambda_D} \right) \right] \\
&= Z_\alpha e^{-r/\lambda_D} \left(1 + \frac{r}{\lambda_D} \right)
\end{align}

\newpage

\section{Evolution of the charge-number coupling at various lithium salt fractions}\label{si:sec:gZN}

\begin {figure*}[bth]
\centering
\includegraphics[width=5in]{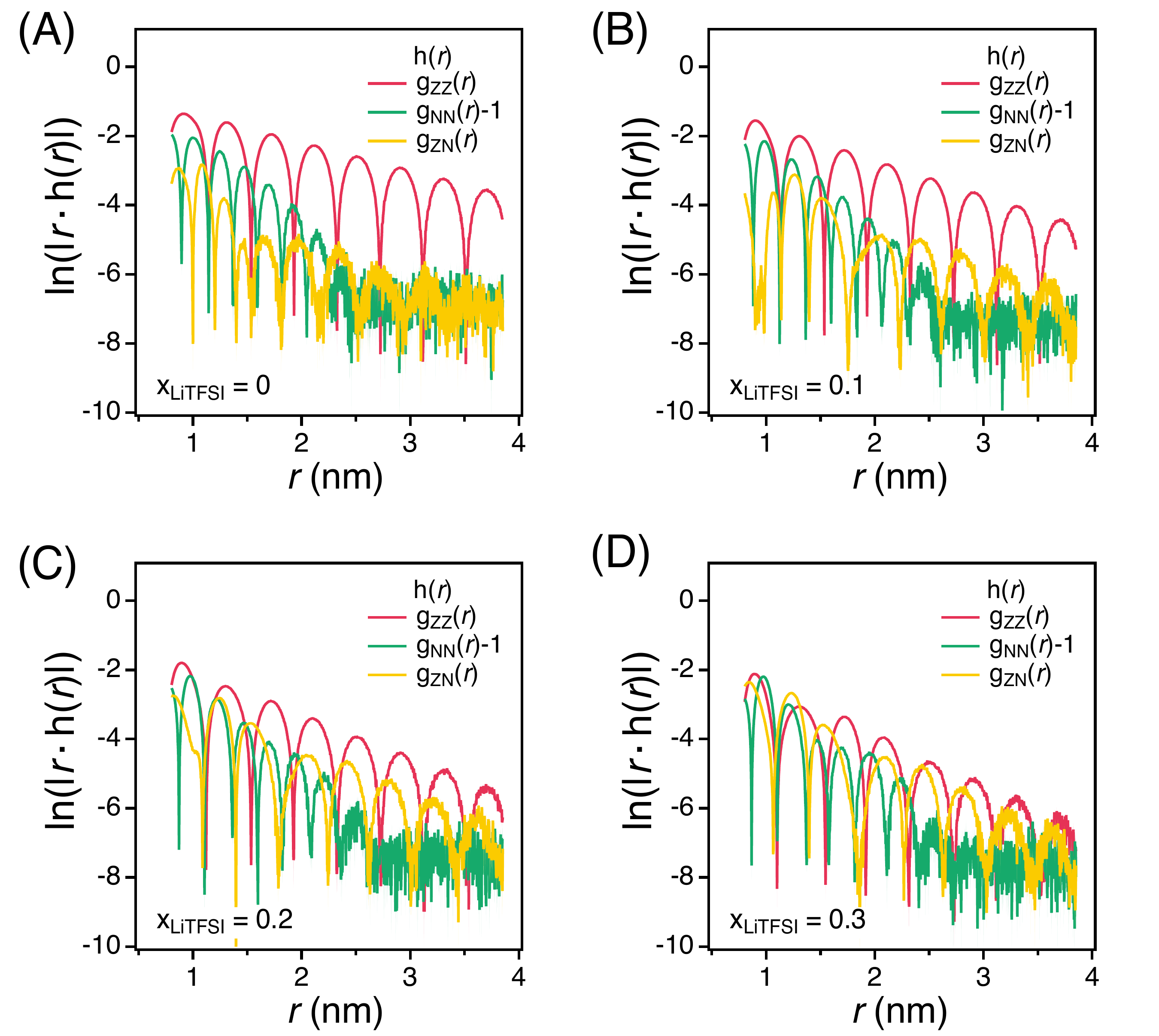}
\caption{Comparison of three structural correlation functions at various lithium salt fractions $\xli$. Each panel presents the results at each $\xli$, and different colors denote different correlation functions.}
\label{si:fig:corr}
\end{figure*}

\newpage

\section{Structural correlation function fits: Single vs. three decay modes}\label{si:sec:fit}

\begin {figure*}[bth]
\centering
\includegraphics[width=4.2in]{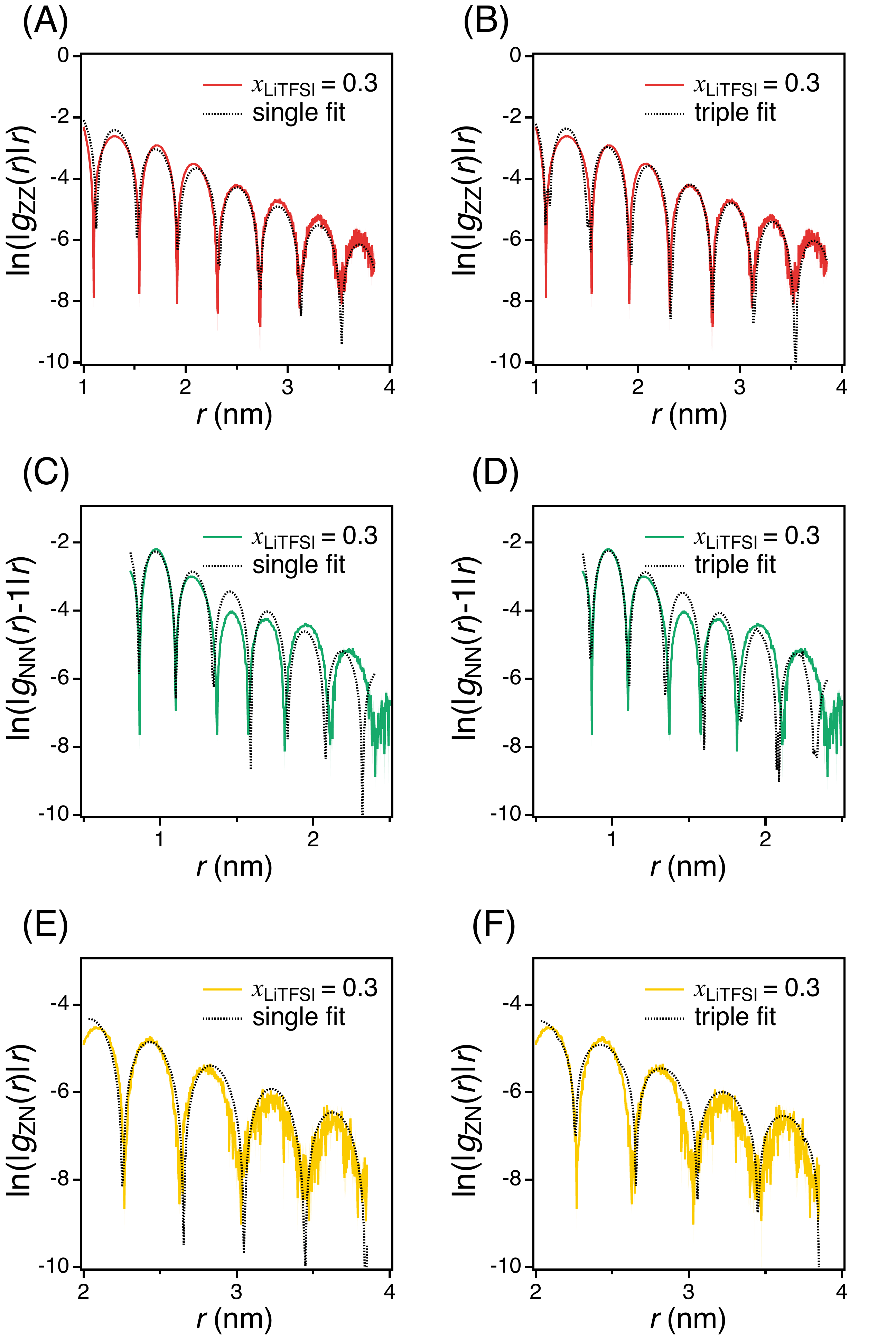}
\caption{Comparison of fitting qualities for structural correlation functions at lithium salt fraction $\xli$ = 0.3. The left column presents the fitting results obtained using a single Yukawa-type function, whereas the right column shows the results obtained using a linear combination of three Yukawa-type functions. (A, B) $g_{ZZ}(r)$. (C, D) $g_{NN}(r)$. (E, F) $g_{ZN}(r)$.}
\label{si:fig:fit}
\end{figure*}

\newpage

\section{Lithium-ion solvation environment by \ce{TFSI} anions}\label{si:sec:solv}

% Lithium coordination
\begin {figure*}[bth]
\centering
\includegraphics[width=6in]{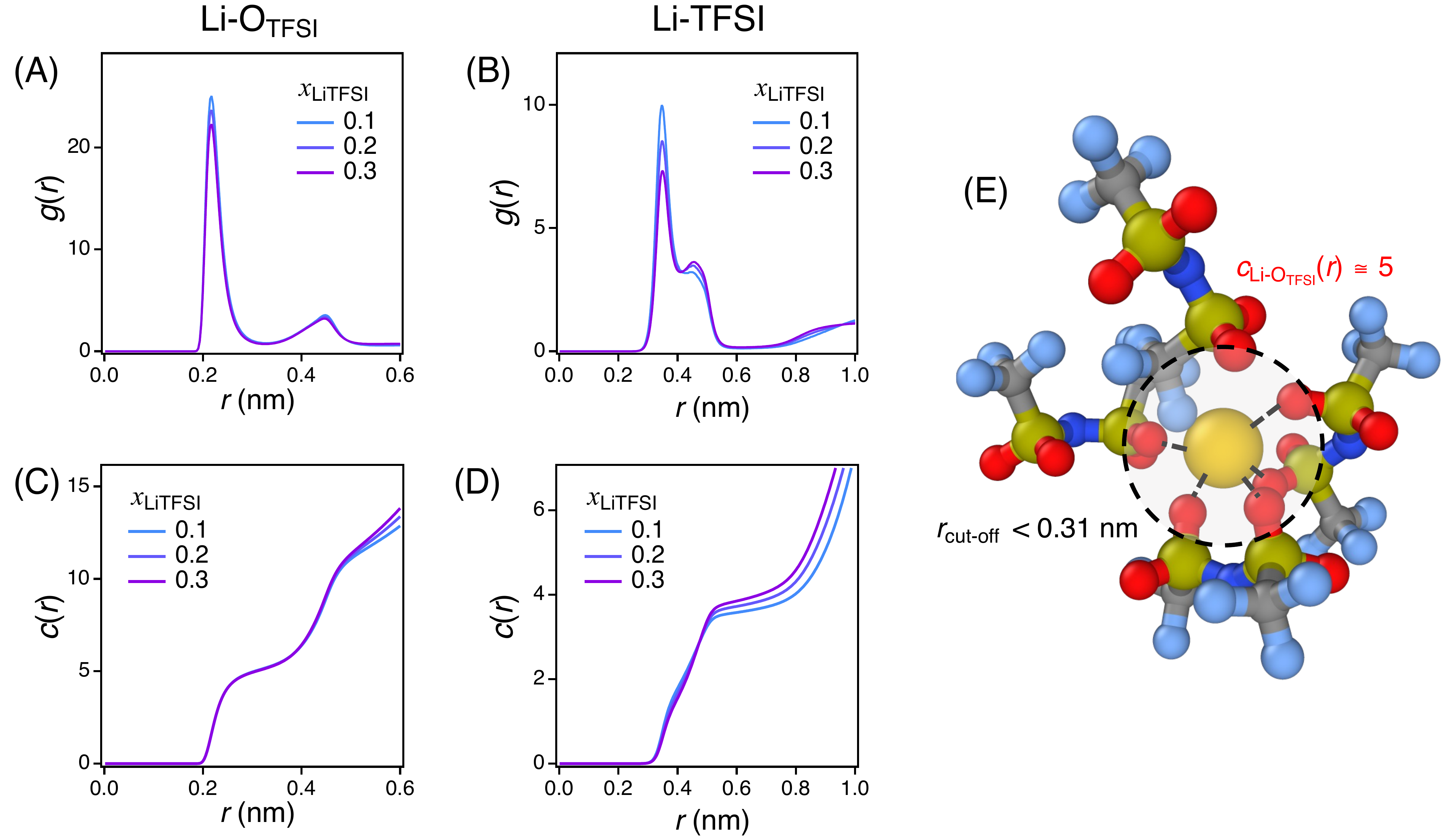}
\caption{Lithium cation solvation environment at various $\litfsi$ doping fractions $\xli$. (A, B) The radial $g(r)$ and (C, D) the cumulative $c(r)$ distributions of $\tfsi$ anions around a $\li$ ion at various doping fractions $\xli$. (A) and (C) represent $\li$ ion solvation structure with the oxygen atoms of $\tfsi$ ions, while (B) and (D) represent that with the center of mass of $\tfsi$ ions. (E) Simulation snapshot of lithium-ion coordination environment at $\xli=0.2$. Here, $c_{\ce{Li}-\ce{O}_{\ce{TFSI}}}(r=0.31~\ce{nm})\approx5$.} 
\label{si:fig:Li_coord}
\end{figure*}

\newpage

\section{Asymmetric Li-TFSI clusters}\label{si:sec:cluster}

\begin {figure*}[h]
\centering
\includegraphics[width=6.in]{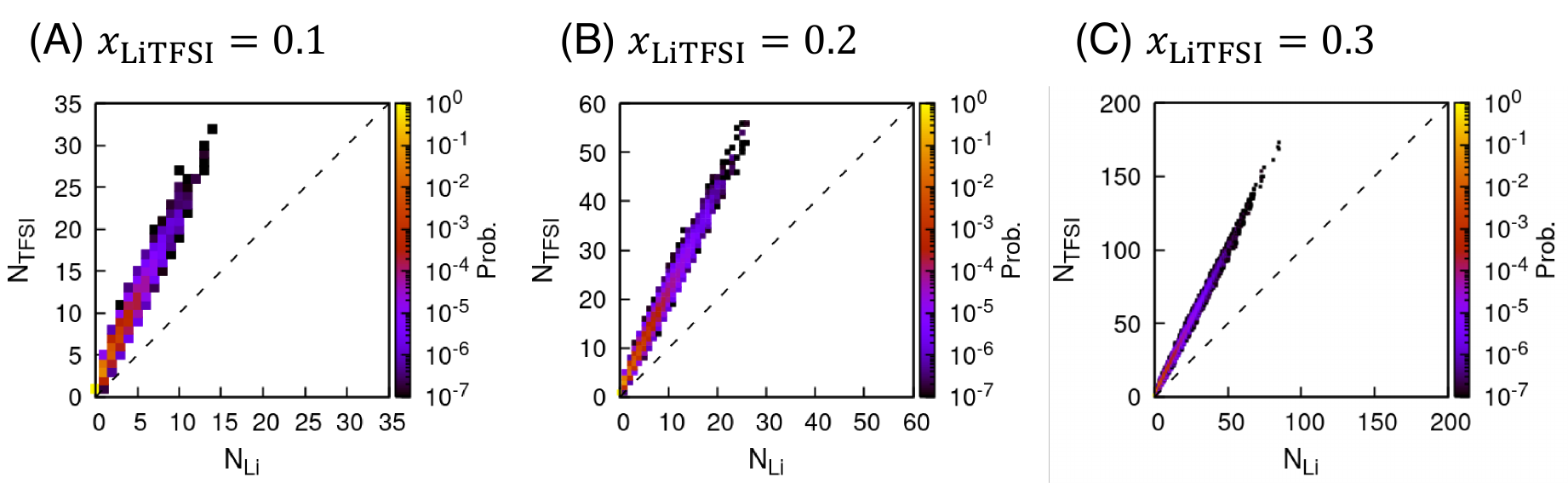}
\caption{
Results of the \ce{Li-TFSI} cluster analysis performed using the TRAVIS software~\cite{brehm2011travis,brehm2020travis}. The distance-based criterion $r_{min}$ (Table~\ref{si:tab:firstshell}) was used to determine whether $\li$ and oxygen atoms of $\tfsi$ belong the same cluster. The panels show the probability distribution of \ce{Li-TFSI} clusters of $N_{\ce{Li}}$ number of \ce{Li}$^+$ ions and $N_{\ce{TFSI}}$ number of \ce{TFSI}$^-$ ions at various $\xli$: (A) $\xli = 0.1$, (B) $\xli = 0.2$, and (C) $\xli = 0.3$. The color scale denotes the probability of observing a \ce{Li-TFSI} cluster with composition ($N_{\ce{Li}}$, $N_{\ce{TFSI}}$) among all clusters. The dashed lines indicate the case of $N_{\ce{Li}} = N_{\ce{TFSI}}$.
}
\label{si:fig:cluster_lian}
\end{figure*}

\newpage

\section{Mean-squared displacement: The self, distinct, and cross terms}\label{si:sec:diff}
\begin {figure*}[]
\includegraphics [width=5in] {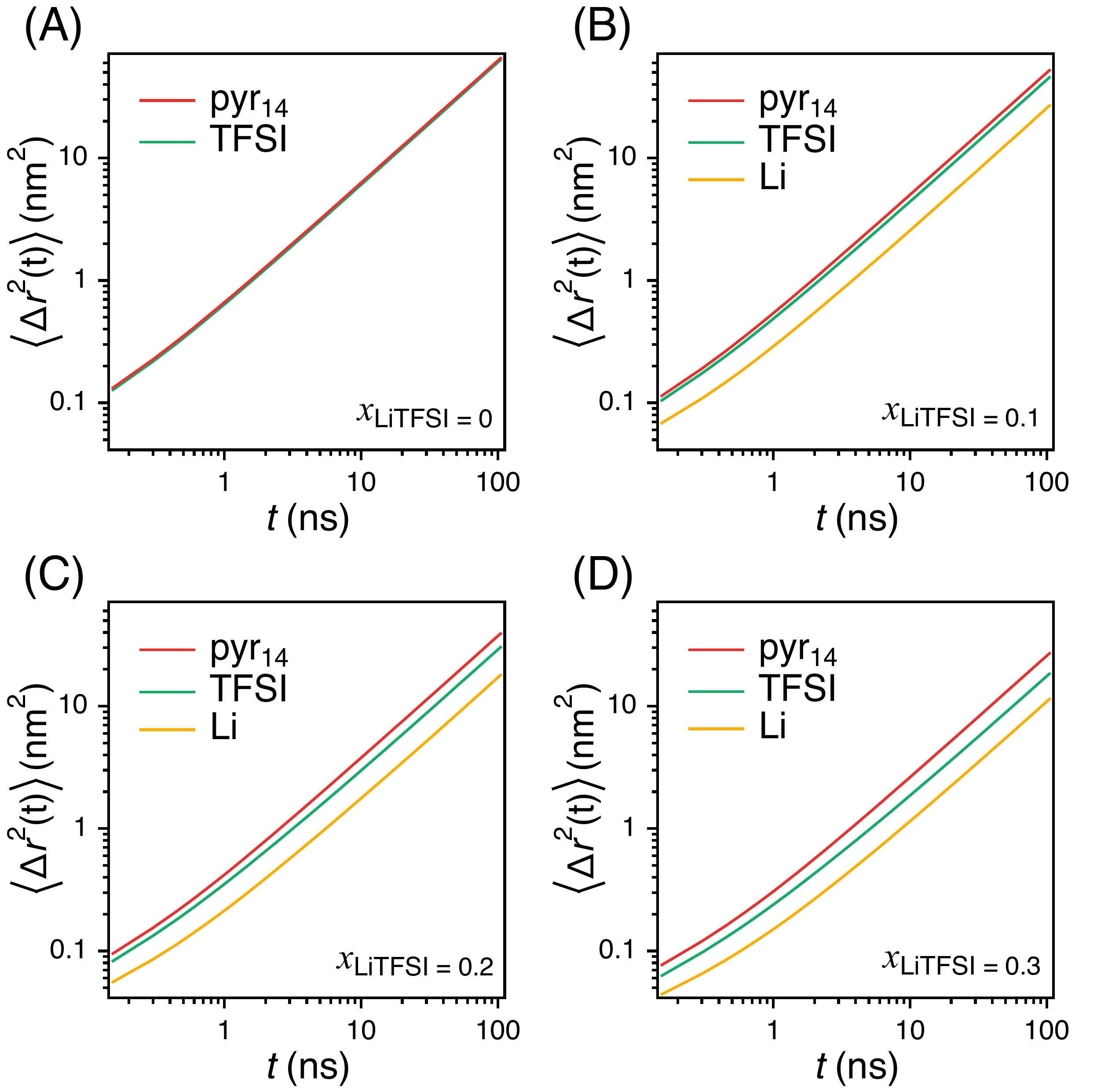}
\caption{The self part of mean-squared displacements of ions ($(\Delta \vec{r}(t))^2$, Eq.~\ref{si:eq:Dself}) at various $\xli$.}
\label{si:fig:msd}
\end{figure*}

\begin {figure*}[h]
\includegraphics [width=5in] {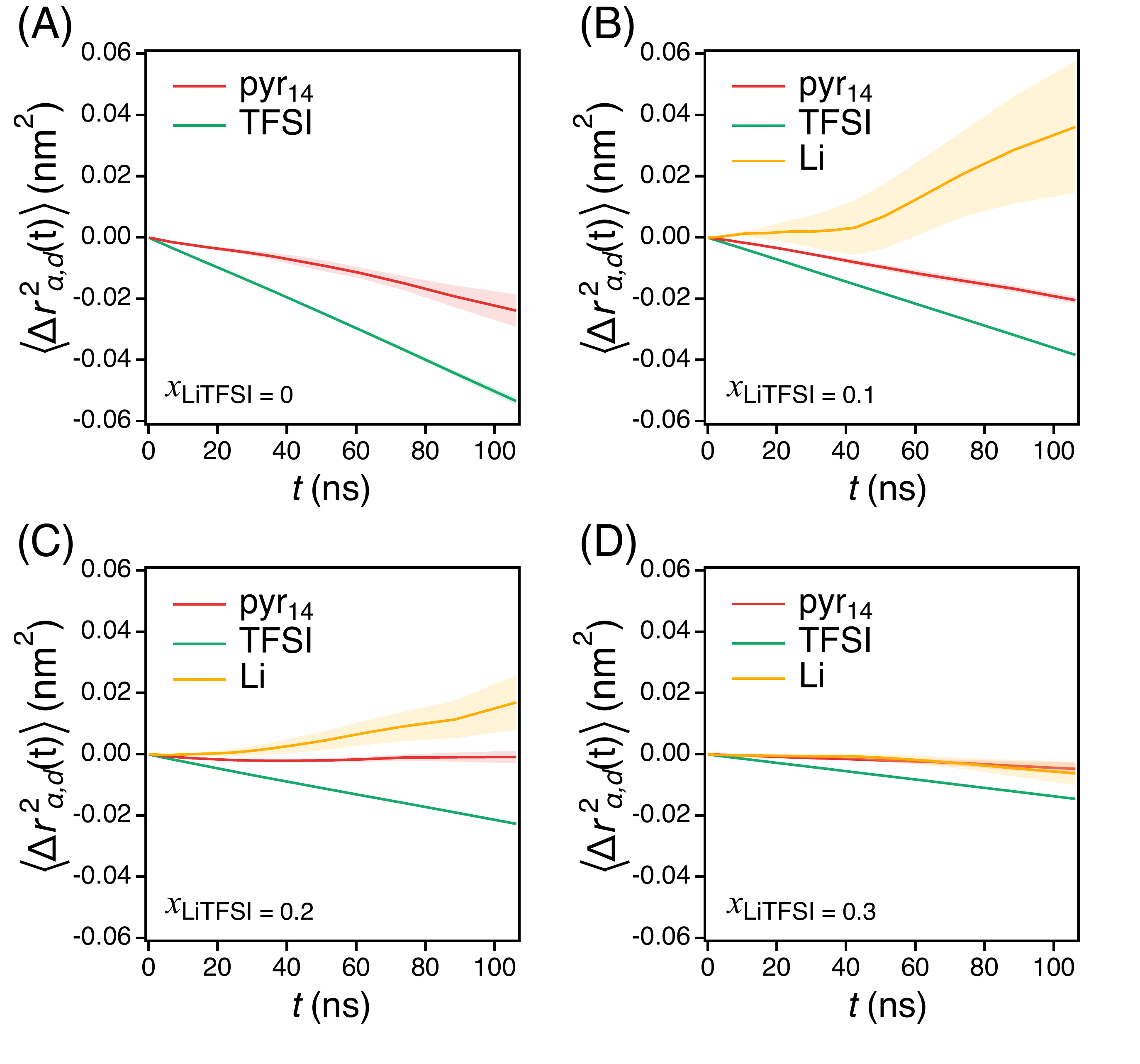}
\caption{The distinct part of mean-squared displacements (Eq.~\ref{si:eq:Ddist}) between the same ionic species at various $\xli$.}
\label{si:fig:dist_same}
\end{figure*}

\begin {figure*}[b]
\includegraphics [width=5in] {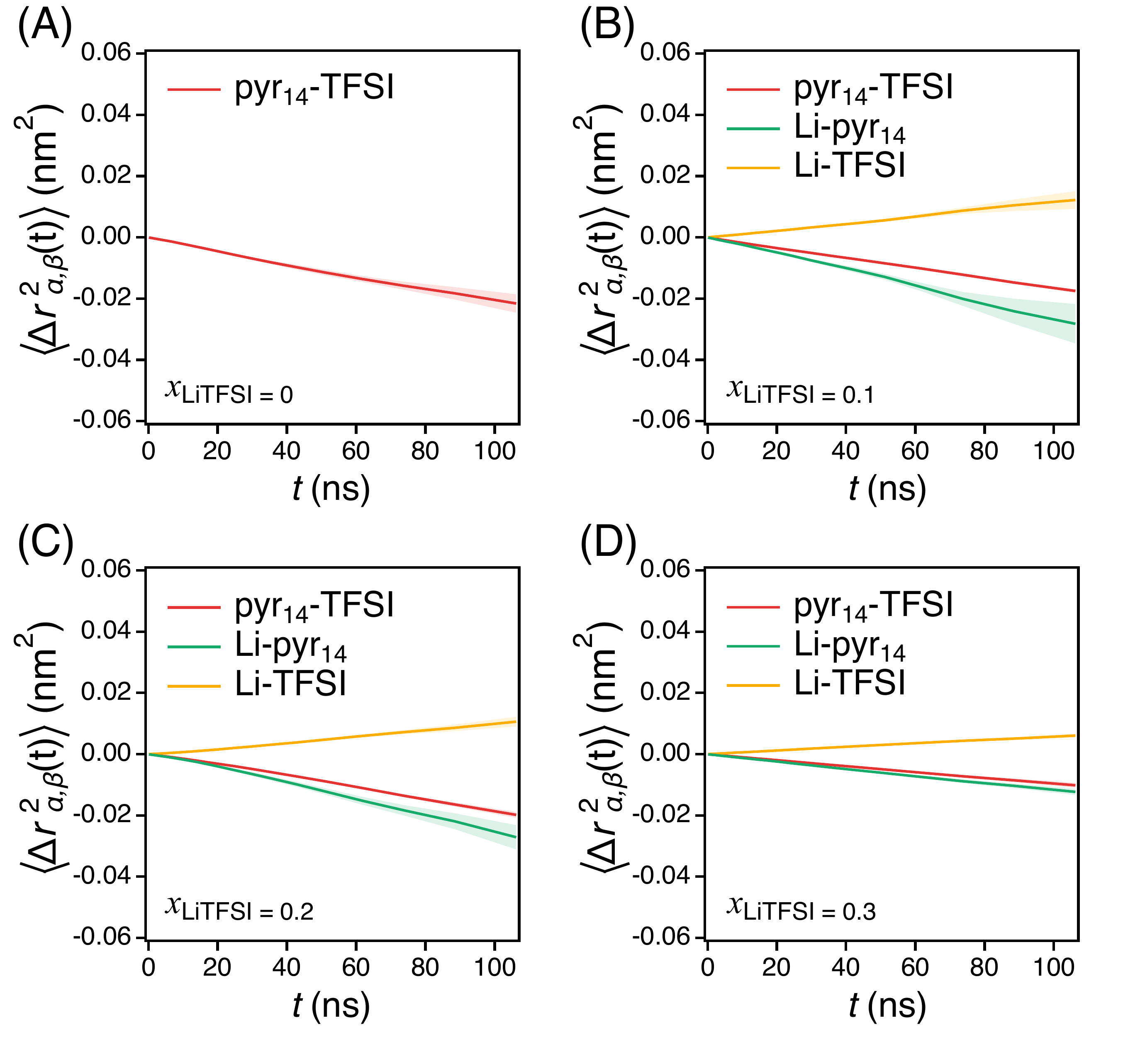}
\caption{The cross part of mean-squared displacements (Eq.~\ref{si:eq:Dcross}) of ions $\alpha$ and $\beta$ at various $\xli$.}
\label{si:fig:dist_cross}
\end{figure*}

We quantify ion diffusion using self- and cross-terms of mean-squared displacements (MSDs). In general, one can define the displacements product $\Delta r^{\alpha\beta}_{ij}$ as follows;
\begin{eqnarray}
\Delta r^{\alpha\beta}_{ij} \equiv \langle \Delta \vec{r}_j^{\alpha}(t) 
\cdot \Delta \vec{r}_i^{\beta}(t) \rangle,
\end{eqnarray}
where $\Delta \vec{r}_i^{\alpha}(t)~(=\vec{r}_i^{\alpha}(t)- \vec{r}_i^{\alpha}(0))$ denotes the displacement of the $i^{th}$ ion of species $\alpha$ during time $t$. Then, we can classify the $\Delta r^{\alpha\beta}_{ij}$ into three terms. Firstly, the conventional MSD $\langle\Delta r^2_{\alpha,s}(t)\rangle$ of species $\alpha$ is the self-term of $\Delta r^{\alpha\beta}_{ij}$ as follows,
\begin{eqnarray}\label{eq:dr_self}
\langle\Delta r^2_{\alpha,s}(t)\rangle=\frac{1}{N_{\alpha}} \sum^{}_{i\in\alpha}\Delta r^{\alpha\alpha}_{ii}.
\end{eqnarray}
The cross terms have two contributions from the same species $\langle\Delta r^2_{\alpha,d}(t)\rangle$ or different species $\langle\Delta r^2_{\alpha,\beta}(t)\rangle$:
\begin{eqnarray}\label{eq:dr_d1}
\langle\Delta r^2_{\alpha,d}(t)\rangle=\frac{2}{N_{\alpha}(N_{\alpha}-1)}\sum^{}_{i\in\alpha}\sum_{j\in\alpha, i\neq j} \Delta r^{\alpha\alpha}_{ij},
\end{eqnarray}
\begin{eqnarray}\label{eq:dr_d2}
\langle\Delta r^2_{\alpha,\beta}(t)\rangle=\frac{1}{N_{\alpha}N_{\beta}}\sum^{}_{i\in\alpha}\sum_{j\in\beta} \Delta r^{\alpha\beta}_{ij}.
\end{eqnarray}
One can compute the diffusion coefficient of the ionic species $\alpha$ using $\langle\Delta r^2_{\alpha,s}(t)\rangle$ in the long-time limit where the MSD grows linearly with time (Fig.~\ref{si:fig:msd}).
\begin{eqnarray}\label{si:eq:Dself}
D_{\alpha}= \lim_{t\to\infty} \frac{\langle\Delta r^2_{\alpha,s}(t)\rangle}{6t}.
\label{eq:diff}
\end{eqnarray}
Similarly, one can compute the diffusion coefficient of the distinct part of the same species $\alpha$ (Fig.~\ref{si:fig:dist_same}):
\begin{eqnarray}\label{si:eq:Ddist}
D_{\alpha,d}= \lim_{t\to\infty} \frac{\langle\Delta r^2_{\alpha,d}(t)\rangle}{6t},
\end{eqnarray}
and the cross part of the different species $\alpha$ and $\beta$ (Fig.~\ref{si:fig:dist_cross}):
\begin{eqnarray}\label{si:eq:Dcross}
D_{\alpha,\beta}= \lim_{t\to\infty} \frac{\langle\Delta r^2_{\alpha,\beta}(t)\rangle}{6t}.
\end{eqnarray}
All the MSDs grow linearly in time in the simulation time window.

\newpage

\section{Correlated ion motions and contributions to the ionic conductivity}\label{si:sec:cond}

\begin {figure*}[h]
\includegraphics [width=4.in] {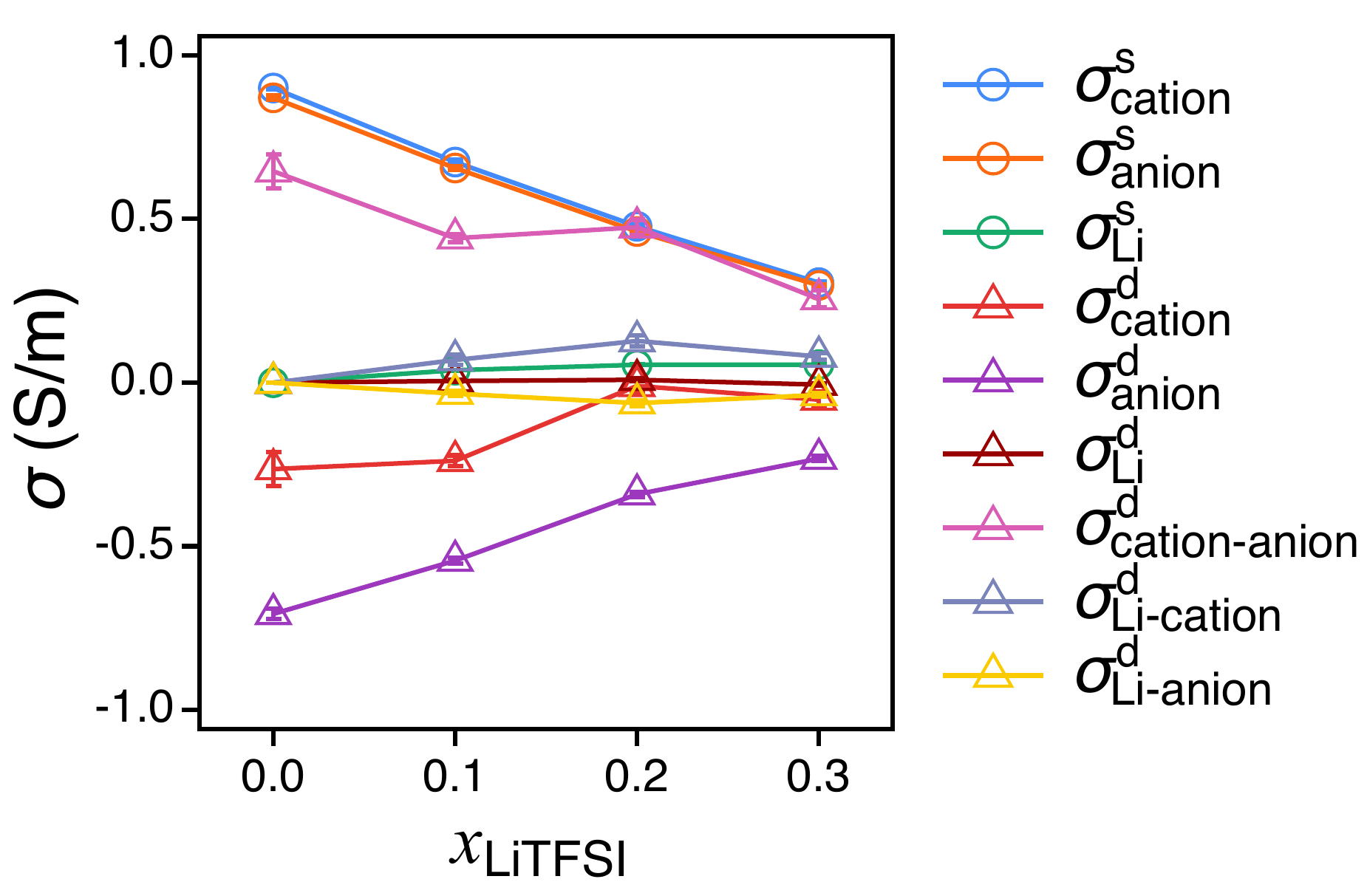}
\caption{Ion conductivity and its contributions from correlated motions at various $\xli$.}
\label{si:fig:contribution}
\end{figure*}

\textbf{Contributions to ionic conductivity.} One can decompose the contributions to the total ionic conductivity as for the ion diffusion: The self contribution $\sigma_{\alpha}^s$, the distinct contribution of the same species $\sigma_{\alpha}^d$, and the cross contribution of the different species $\sigma_{\alpha-\beta}^d$. Their long-time limits are computed using the diffusion coefficients. The contributions at various $\xli$ are displayed in Fig.~\ref{si:fig:contribution}.

\begin{gather*} 
\sigma_{\alpha}^s= \lim_{t\to\infty}\frac{Z^2_\alpha}{6Vtk_BT} \sum^{N_\alpha}_{i\in\alpha}\Delta\vec{r}^{\alpha\alpha}_{ii}=\frac{N_{\alpha}Z^2_\alpha}{Vk_BT}D_{\alpha}, \\ \nonumber
\sigma_{\alpha}^d= \lim_{t\to\infty}\frac{Z^2_\alpha}{6Vtk_BT} \sum^{N_\alpha}_{i\in\alpha} \sum^{N_\alpha}_{j\in\alpha, i\neq j} \Delta\vec{r}^{\alpha\alpha}_{ij}=\frac{N_{\alpha}(N_{\alpha}-1)Z^2_\alpha}{Vk_BT}D_{\alpha,d}, \\ \nonumber 
\sigma^d_{\alpha-\beta}=\lim_{t\to\infty}\frac{Z_\alpha Z_\beta}{6Vtk_BT} \sum^{N_\alpha}_{i\in\alpha} \sum^{N_\beta}_{j\in\beta} \Delta\vec{r}^{\alpha\beta}_{ij}=\frac{N_{\alpha}N_{\beta}Z_\alpha Z_\beta}{Vtk_BT}D_{\alpha,\beta} \text{~for~} \alpha \neq \beta.
\end{gather*} 

\clearpage

\begin {figure}[]
\centering
\includegraphics[width=5.5in]{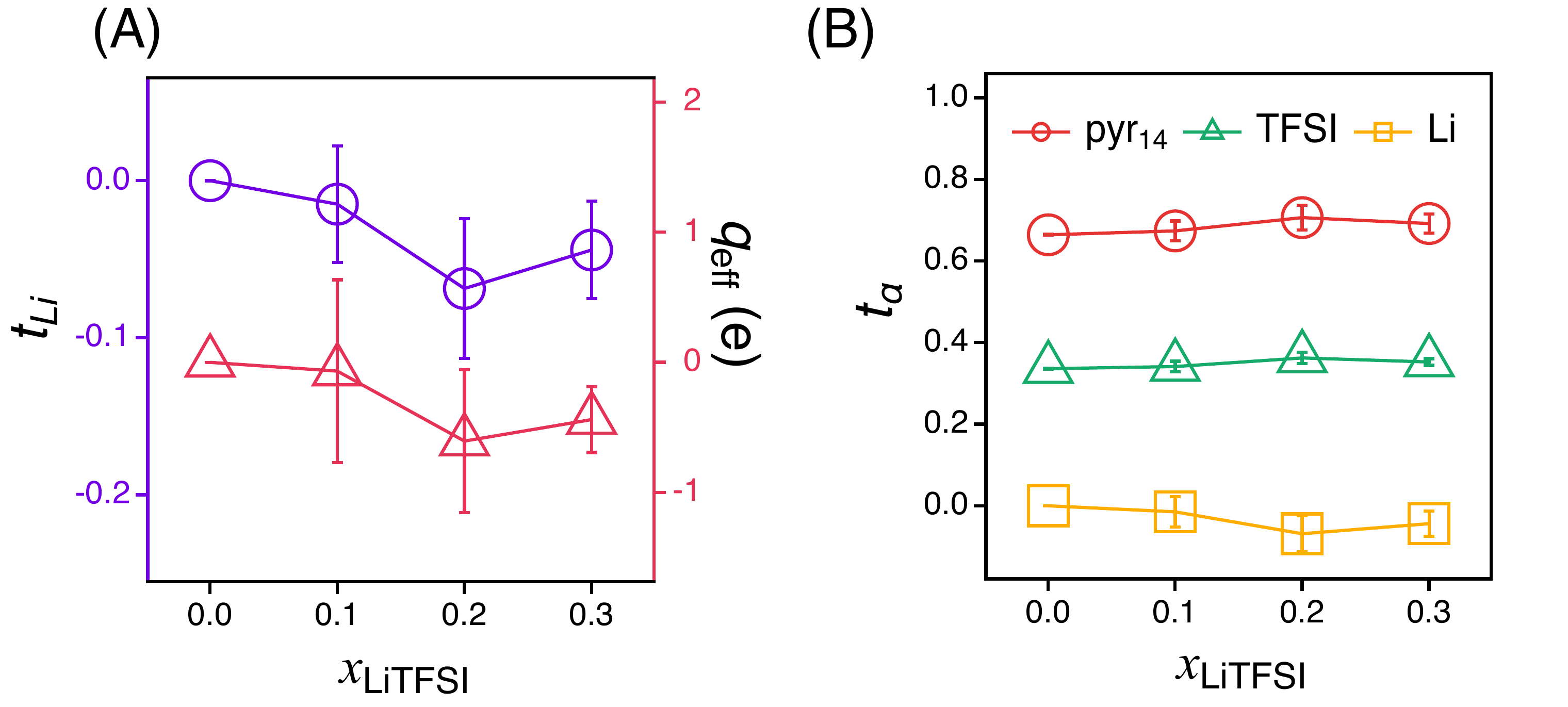}
\caption{Simulation results of ion transport at various $\litfsi$ doping fractions $\xli$. (A) The lithium transference number $t_{\ce{Li}}$ and its effective charge $q_{\text{eff}}$ (Eq.~\ref{si:eq:qeff}). (B) The transference number $t_{\alpha}$ of all three constituent ions. Here, $t_{\ce{Li}}=0$ at $\xli=0$ by definition.
} 
\label{si:fig:cond}
\end{figure}

To highlight the effect of correlation in ion motion, the effective charge $q_{\mathrm{eff}}$ transported by lithium can be defined using the actual and NE transference numbers~\cite{Molinari2019,McEldrew2021b}:
\begin{equation}\label{si:eq:qeff}
q_{\mathrm{eff}} = \mathrm{sign}(t_{\ce{Li}}) \sqrt{ \frac{ |t_{\ce{Li}}| }{ t_{\ce{Li}}^{\mathrm{NE}} } }.
\end{equation}
If $t_{\ce{Li}} = t_{\ce{Li}}^{\mathrm{NE}}$, then $q_{\mathrm{eff}} = 1$, indicating no contribution of correlations in ion motion. By definition, $t_{\ce{Li}}^{\mathrm{NE}}>0$ since it is derived solely from the self-diffusion coefficients. The sign function accounts for possible negative transference numbers, where lithium effectively migrates in the direction opposite to the net ionic current.

\clearpage

\section{Ion pair dynamics $H(t)$}\label{si:sec:pairdyn}

\begin {figure*}[h]
\centering
\includegraphics[width=6in]{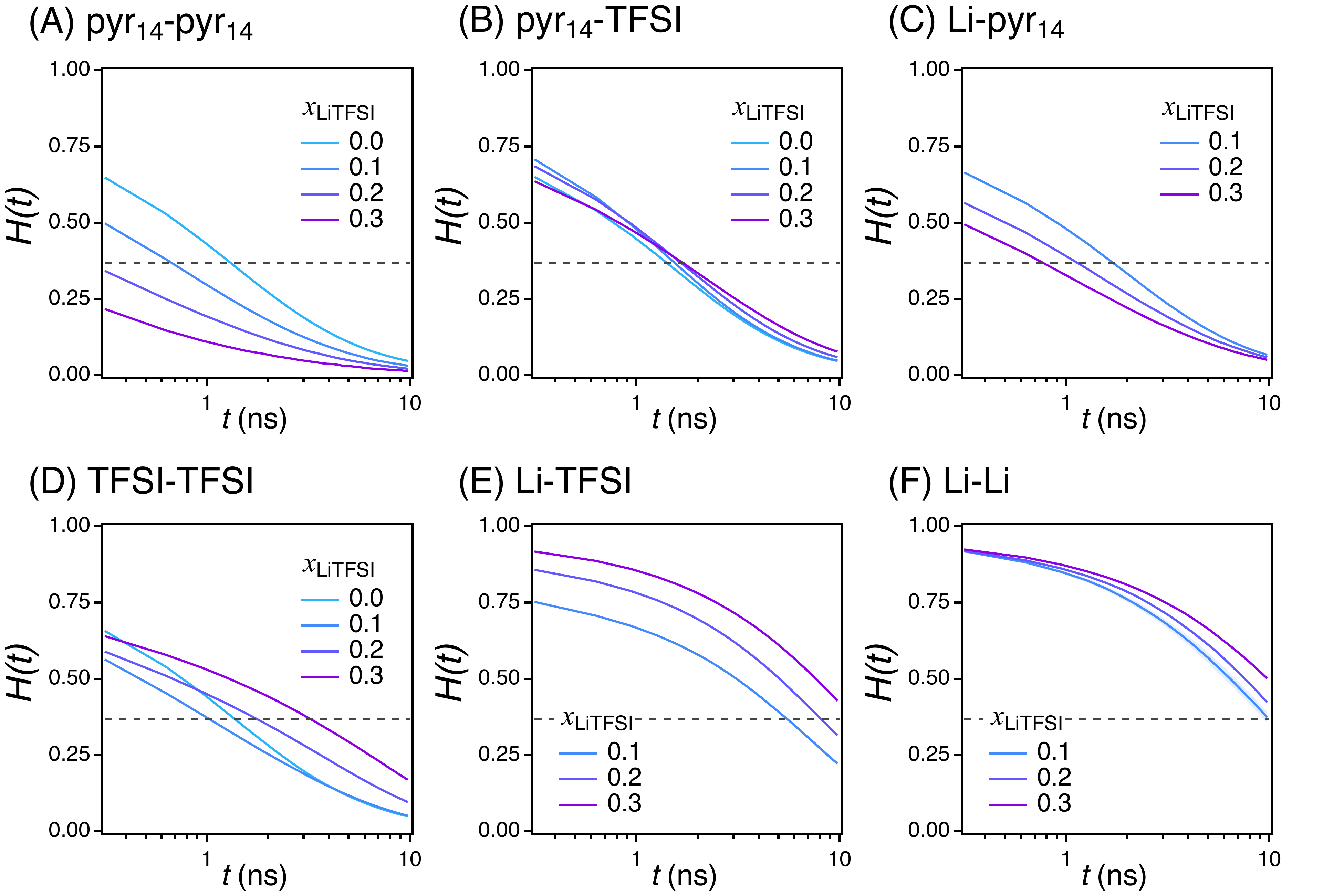}
\caption{Ion pair dynamics ($H(t)$, Eq.~\ref{eq:pairdyn}) at various $\xli$ concentrations. Here, ion pairs are identified using $\lambda_Z$ as a distance criterion.}
\label{si:fig:pairdyn_lz}
\end{figure*}

\begin {figure*}[h]
\centering
\includegraphics[width=6in]{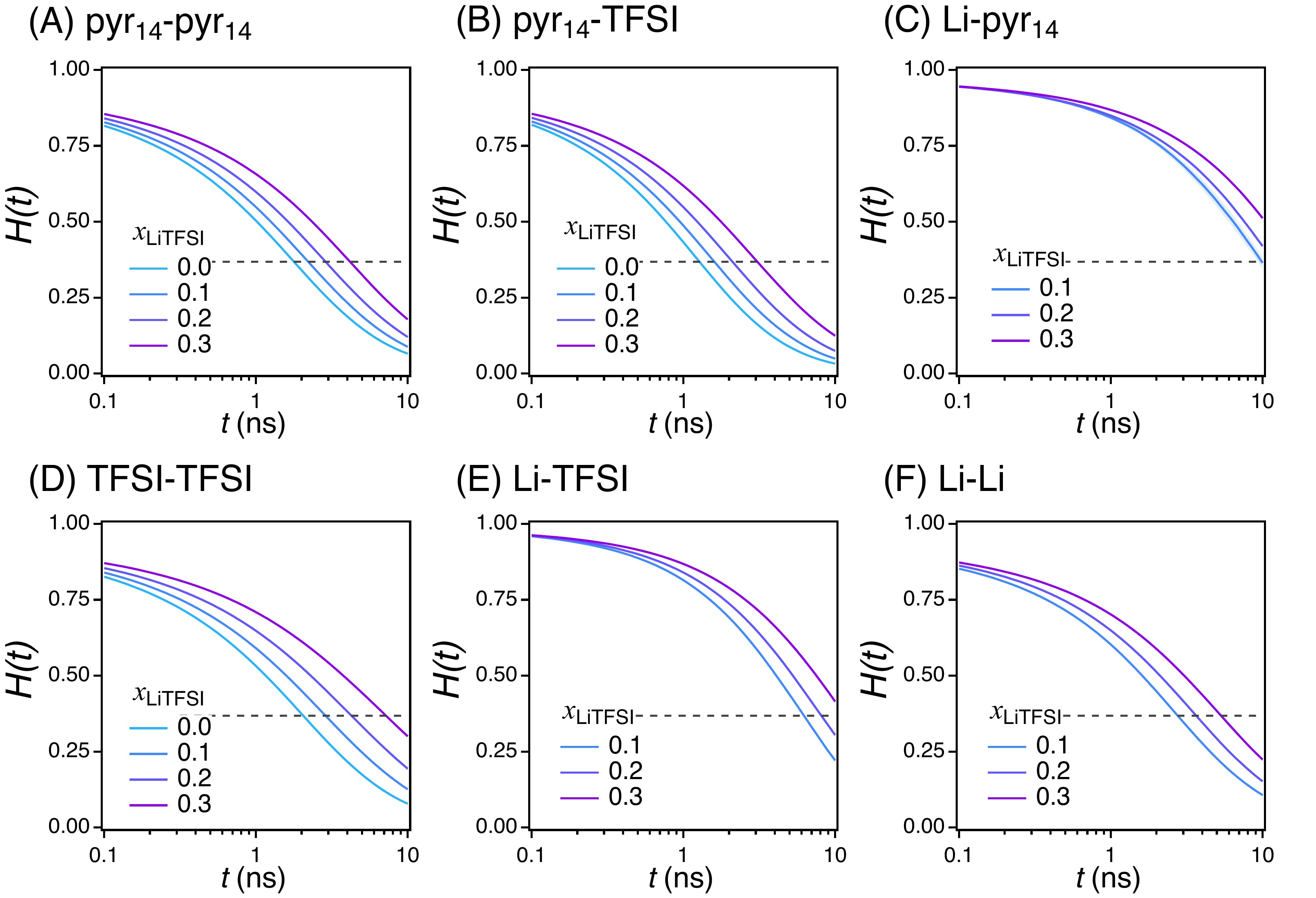}
\caption{Ion pair dynamics ($H(t)$, Eq.~\ref{eq:pairdyn}) at various $\xli$ concentrations. Here, ion pairs are identified using $r_{min}$ (Table~\ref{si:tab:firstshell}) as a distance criterion.}
\label{si:fig:pairdyn_rmin}
\end{figure*}

\begin {figure*}[h]
\centering
\includegraphics[width=6in]{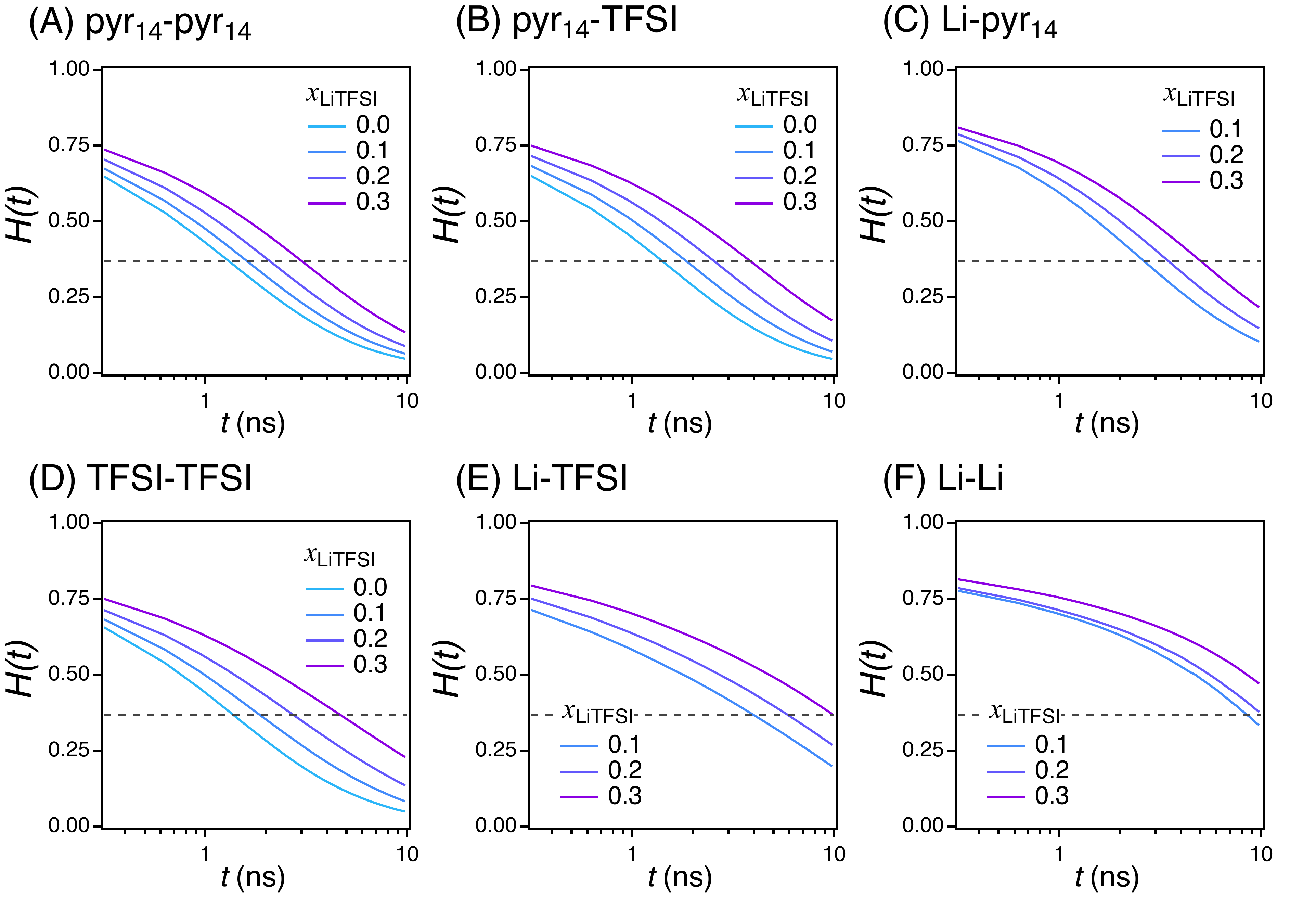}
\caption{Ion pair dynamics ($H(t)$, Eq.~\ref{eq:pairdyn}) at various $\xli$ concentrations. Here, ion pairs are identified using $\lambda_Z$ of the pure $\ce{pyr_{14}TFSI}$ at $\xli=0$ as a distance criterion.}
\label{si:fig:pairdyn_fixl}
\end{figure*}

\begin{table}[!htbp]
\begin{ruledtabular}
\begin{tabular}{l|lll|lll}
 & $\ce{pyr14}-\ce{pyr14}$ & & & $\ce{TFSI}-\ce{TFSI}$ & &  \\ \hline
$\xli$ & $\tau_{\text{H}}$ & $\beta_{\text{H}}~~~~~~$ & $\langle\tau_H\rangle$~~~~~~~~~~ & $\tau_{\text{H}}$ & $\beta_{\text{H}}$~~~~~~ & $\langle\tau_H\rangle$~~~~~~~~~~\\ 
0.0 & 1.31 & 0.58 & 2.05 & 1.38 & 0.59 & 2.15 \\ 
0.1 & 0.66 & 0.48 & 1.45 & 1.02 & 0.48 & 2.21 \\
0.2 & 0.26 & 0.37 & 1.07 & 1.58 & 0.43 & 4.34 \\
0.3 & 0.07 & 0.29 & 0.71 & 2.87 & 0.40 & 9.32 \\ \hline \hline
& $\ce{pyr14}-\ce{TFSI}$ & & & $\ce{Li}-\ce{TFSI}$ & & \\ \hline
$\xli$ & $\tau_{\text{H}}$ & $\beta_{\text{H}}$~~~~~~ & $\langle\tau_H\rangle$~~~~~~~~~~ & $\tau_{\text{H}}$ & $\beta_{\text{H}}$~~~~~~ & $\langle\tau_H\rangle$~~~~~~~~~~\\ 
0.0 & 1.38 & 0.59 & 2.12 & - & - & - \\ 
0.1 & 1.59 & 0.64 & 2.21 & 5.4  & 0.51 & 10.5 \\
0.2 & 1.64 & 0.60 & 2.47 & 9.1  & 0.58 & 14.4 \\
0.3 & 1.63 & 0.53 & 2.98 & 14.0 & 0.70 & 17.8 \\ \hline \hline
& $\ce{Li}-\ce{TFSI}$ & & & $\ce{Li}-\ce{Li}$ & & \\ \hline
$\xli$ & $\tau_{\text{H}}$ & $\beta_{\text{H}}$~~~~~~ & $\langle\tau_H\rangle$~~~~~~~~~~ & $\tau_{\text{H}}$ & $\beta_{\text{H}}$~~~~~~ & $\langle\tau_H\rangle$~~~~~~~~~~\\ 
0.1 & 1.67 & 0.57 & 2.67 & 10.7 & 0.74 & 13.0 \\
0.2 & 1.09 & 0.48 & 2.39 & 12.5 & 0.75 & 14.8 \\
0.3 & 0.76 & 0.43 & 2.09 & 20.0 & 0.65 & 27.2 \\
\end{tabular}
\caption{Fit results for $H(t)$ with the distance criterion $\lambda_Z$ (Fig.~\ref{si:fig:pairdyn_lz}). Here, $H(t)$ is fitted with a stretched exponential function $\text{exp}(-(t/\tau_{\text{H}})^{\beta_{\text{H}}})$. Mean lifetime is estimated as $\langle\tau_H\rangle=\tau_{\text{H}}/\beta_{\text{H}}\Gamma(1/\beta_{\text{H}})$.}
\label{si:tab:hfit_lz}
\end{ruledtabular}
\end{table}

\begin{table}[!htbp]
\begin{ruledtabular}
\begin{tabular}{l|lll|lll}
 & $\ce{pyr14}-\ce{pyr14}$ & & & $\ce{TFSI}-\ce{TFSI}$ & &  \\ \hline
$\xli$ & $\tau_{\text{H}}$ & $\beta_{\text{H}}~~~~~~$ & $\langle\tau_H\rangle$~~~~~~~~~~ & $\tau_{\text{H}}$ & $\beta_{\text{H}}$~~~~~~ & $\langle\tau_H\rangle$~~~~~~~~~~\\ 
0.0 & 1.72 & 0.57 & 2.80 & 1.92 & 0.56 & 3.16 \\ 
0.1 & 2.04 & 0.55 & 3.45 & 2.64 & 0.52 & 4.91 \\
0.2 & 2.90 & 0.63 & 4.13 & 4.27 & 0.59 & 6.65 \\
0.3 & 4.16 & 0.63 & 5.91 & 7.30 & 0.55 & 12.5 \\ \hline \hline
& $\ce{pyr14}-\ce{TFSI}$ & & & $\ce{Li}-\ce{TFSI}$ & & \\ \hline
$\xli$ & $\tau_{\text{H}}$ & $\beta_{\text{H}}$~~~~~~ & $\langle\tau_H\rangle$~~~~~~~~~~ & $\tau_{\text{H}}$ & $\beta_{\text{H}}$~~~~~~ & $\langle\tau_H\rangle$~~~~~~~~~~\\ 
0.0 & 1.30 & 0.64 & 1.80 & - & - & - \\ 
0.1 & 1.52 & 0.60 & 2.30 & 6.39  & 0.80 & 7.27 \\
0.2 & 2.11 & 0.64 & 2.94 & 8.13  & 0.84 & 8.88 \\
0.3 & 3.01 & 0.64 & 4.27 & 12.00 & 0.80 & 13.5 \\ \hline \hline
& $\ce{Li}-\ce{TFSI}$ & & & $\ce{Li}-\ce{Li}$ & & \\ \hline
$\xli$ & $\tau_{\text{H}}$ & $\beta_{\text{H}}$~~~~~~ & $\langle\tau_H\rangle$~~~~~~~~~~ & $\tau_{\text{H}}$ & $\beta_{\text{H}}$~~~~~~ & $\langle\tau_H\rangle$~~~~~~~~~~\\ 
0.1 & 6.39 & 0.80 & 7.27 & 11.2 & 0.79 & 14.1 \\
0.2 & 8.13 & 0.85 & 8.88 & 12.5 & 0.75 & 15.0 \\
0.3 & 12.0 & 0.80 & 13.5 & 20.0 & 0.67 & 26.2 \\
\end{tabular}
\caption{Fit results for $H(t)$ with the distance criterion $r_{min}$ (Fig.~\ref{si:fig:pairdyn_rmin})\textemdash the size of the first solvation shell (Table~\ref{si:tab:firstshell}). Here, $H(t)$ is fitted with a stretched exponential function $\text{exp}(-(t/\tau_{\text{H}})^{\beta_{\text{H}}})$. The mean lifetime is estimated as $\langle\tau\rangle=\tau_{\text{H}}/\beta_{\text{H}}\Gamma(1/\beta_{\text{H}})$. The values of $r_{min}$ are listed in Fig.~\ref{si:fig:rdf}.}
\label{si:tab:hfit_rmin}
\end{ruledtabular}
\end{table}

\clearpage

%\bibliography{SI}
